\documentclass[11pt,a4paper,english]{article}
\usepackage{babel}
\usepackage{epsfig}
\usepackage{times}
\usepackage{amssymb}
\usepackage[comma,authoryear,sort,longnamesfirst]{natbib}

\renewcommand{\baselinestretch}{1}
\begin{document}

\title{
\vspace{-15mm}
\begin{flushright}
    \footnotesize{Accepted in the Journal of Statistical Software, January 2010}
\end{flushright}
Measures of Analysis of Time Series ({\tt MATS}): A {\tt MATLAB} Toolkit for Computation of Multiple Measures on Time Series Data Bases}

\author{Dimitris Kugiumtzis, Alkiviadis Tsimpiris \\
{\em\small Department of Mathematical, Physical and Computational
Sciences} \\
{\em\small Faculty of Engineering, Aristotle University of
Thessaloniki} \\
{\em\small Thessaloniki 54124, Greece}
}

\date{}
\maketitle

\bigskip

\begin{flushleft}
    {\em Corresponding author} \\
    Dimitris Kugiumtzis \\
    Department of Mathematical, Physical and Computational Sciences,\\
    Faculty of Engineering, Aristotle University of Thessaloniki, \\
    Thessaloniki 54124, Greece\\
    e-mail: dkugiu@gen.auth.gr\\
    url: http://users.auth.gr/dkugiu \\
    fax: +30 2310995958
\end{flushleft}

\begin{flushleft}
    {\em Running title} \\
    Optimal Measures and Preictal States
\end{flushleft}

\bibliographystyle{apa}

\begin{abstract}
In many applications, such as physiology and finance, large time series data bases are to be analyzed requiring the computation of linear, nonlinear and other measures. Such measures have been developed and implemented in commercial and freeware softwares rather selectively and independently. The Measures of Analysis of Time Series ({\tt MATS}) {\tt MATLAB} toolkit is designed to handle an arbitrary large set of scalar time series and compute a large variety of measures on them, allowing for the specification of varying measure parameters as well. The variety of options with added facilities for visualization of the results support different settings of time series analysis, such as the detection of dynamics changes in long data records, resampling (surrogate or bootstrap) tests for independence and linearity with various test statistics, and discrimination power of different measures and for different combinations of their parameters. The basic features of {\tt MATS} are presented and the implemented measures are briefly described. The usefulness of {\tt MATS} is illustrated on some empirical examples along with screenshots.
\end{abstract}

\noindent {\em keywords}: time series analysis; data bases; nonlinear dynamics; statistical measures; MATLAB software; change detection; surrogate data.
\newpage

\section{Introduction}

The analysis of time series involves a range of disciplines, from engineering to economics, and its development spans different aspects of the time series, e.g., analysis in the time and frequency domain, and under the assumption of a stochastic or deterministic process \citep{Box94a,Kay88,Tong90,Kantz97}.

Many developed software products include standard methods of time series analysis. Commercial statistical packages, such as {\tt SPSS} \citep{SPSS} and {\tt SAS} \citep{SAS}, have facilities on time series which are basically implementations of the classical Box-Jenkins approach on non-stationary time series. The commercial computational environments {\tt MATLAB} \citep{MATLAB} and {\tt S-PLUS} \citep{SPLUS} provide a number of toolboxes or modules that deal with time series ({\tt MATLAB}: {\tt Financial}, {\tt Econometrics}, {\tt Signal Processing}, {\tt Neural Network} and {\tt Wavelets}, {\tt S-PLUS}: {\tt FinMetrics}, {\tt Wavelet}, {\tt EnvironmentalStats}). Less standard and more advanced tools of time series analysis can be found in a number of commercial stand-alone software packages developed for more specific uses of time series, and in a few freeware packages mostly offering a set of programs rather than an integrated software, such as the {\tt TISEAN} package which includes a comprehensive set of methods of the dynamical system approach \citep{Hegger99b}. Some open-source {\tt MATLAB} toolkits offer selected methods for analyzing an input time series, such as the time series analysis toolbox {\tt TSA} \cite{TSA}.

In many applications, particularly when the system is rather complex as in seismology, climate, finance and physiology, the development of models from data can be a formidable task. There the interest is rather in compressing the information from a time series to some extracted features or measures that can be as simple as descriptive statistics or as complicated as the goodness of fit of a nonlinear model involving many free parameters.
Moreover, the problem at hand may involve a number of time series, e.g., multiple stock indices or consecutive segments from the split of a long physiological signal, such as an electrocardiogram (ECG) or electroencephalogram (EEG). The task of handling these problems is to characterize, discriminate or cluster the time series on the basis of the estimated measures. For example, different types of measures have been applied to financial indices in order to compare international markets at different periods or to each other, such as the NMSE fit \citep{AlvarezDiaz08} and Shannon entropy \citep{Risso09}, and in segmented preictal EEG segments in order to predict the seizure onset
\citep[for a comprehensive collection of measures see e.g.,][]{Mormann05,Kugiumtzis06,Kugiumtzis07}.
It appears that none of the existing commercial or freeware packages are suited to such problems and typically they provide partial facilities, e.g., computing some specific measures, and the user has to write code to complete the analysis.

The {\em measures of analysis of time series} ({\tt MATS}) is developed in {\tt MATLAB} to meet the above-mentioned needs.
The purpose of {\tt MATS} is not only to implement well-known and well-tested measures, but to introduce also other measures, which have been recently presented in the literature and may be useful in capturing properties of the time series. Moreover, {\tt MATS} provides some types of pre-processing of arbitrary long sets of scalar time series. {\tt MATS} exploits the graphical user interface (GUI) of {\tt MATLAB} to establish a user-friendly environment and visualization facilities. In the computation of some methods, functions in the {\tt Signal Processing}, {\tt System Identification} and {\tt Statistics} {\tt MATLAB} toolboxes are called. Measures related to specific toolboxes, such as GARCH models \citep{McCullough98}, neural networks and wavelets, are not implemented in {\tt MATS}. In order to speed up computations when searching for neighboring points, {\tt MATS} uses $k$-$D$ tree data structure built in external files ({\tt .dll} for Windows and {\tt .mexglx} for Linux), which are supported by {\tt MATLAB} versions 6.5 or later.

The structure of the paper is as follows. The implemented facilities on time series pre-processing are presented in Section~\ref{sec:Processing} and in Section~\ref{sec:Measures} the implemented measures are briefly described. Then in Section~\ref{sec:software} the basic features of {\tt MATS} {\tt MATLAB} toolkit are presented and discussed and in Section~\ref{sec:Applications} some representative real-world applications of the software are presented. Finally,
further possible development of {\tt MATS} and improvement of its applicability are discussed in Section~\ref{sec:Discussion}.

\section{Time series pre-processing}
\label{sec:Processing}

The {\tt MATS} toolkit operates on a set of loaded time series, that is referred to as {\em current time series set}. New time series can also be added to the set through pre-processing and resampling operations on the time series in the set.

\subsection{Load and save time series}
\label{subsec:Load}

A flexible loading facility enables the user to select many time series from multiple files at various formats. The selected data files are first stored in a temporary list for format validation. Currently supported formats are the plain text (ascii), excel ({\tt .xls}), {\tt MATLAB} specific ({\tt .mat}) and the European ({\tt .edf}) data format; the latter is used specifically for biological and physical records. Other formats can easily be added. After validation, the loaded data are organized in vector form and a unique name is assigned to each vector, i.e., a unique name for each scalar time series. The loading procedure is completed by adding the imported time series to the current time series set.

The current time series set may expand with time series created by the the pre-processing and resampling operations. It is also possible to select and delete time series from the current time series set. Moreover, any time series in the current set can be saved in files of ascii, {\tt .xls} or {\tt .mat} format.

\subsection{Segmentation and transformation}
\label{subsec:Manipulation}

The pre-processing operations currently implemented are segmentation and transformation of the time series. Other pre-processing operations supported by {\tt MATLAB}, such as filtering, can be easily implemented in the same way.

In some applications, a long time series record is available and the objective is to analyze (e.g., compute measures on) consecutive or overlapping segments from this record.
The facility of {\em time series segmentation} generates consecutive or overlapping segments of a number of selected time series in the current set. The segment length and the sliding step are free parameters set by the user and the user selects also whether the residual data window from the splitting should be in the beginning or at the end of the input time series.

Often real time series are non-stationary and do not have a bell-shaped Gaussian-like distribution. Popular transforms to gain Gaussian marginal distribution are the logarithmic transform and the Box-Cox power transform using an appropriate parameter $\lambda$ \citep{Box64}. For $\lambda=0$, the natural logarithm is taken and if the time series contains negative values the data are first displaced to have minimum 1. To remove slow drifts in the data first differences are taken and to stabilize the variance as well the first differences are taken on the logarithms of the data. The higher-order differences have also been used. Parametric detrending can be done with polynomials of a given degree; the polynomial is first fitted to the data and the residuals comprise the detrended time series. All the above transforms are implemented in the facility of {\em time series transformation}.

When comparing measures across different time series, it is suggested to align the time series in terms of range or marginal distribution, so that any differences of the computed measures on the time series cannot be assigned to differences in the range or shape of the time series amplitudes. Four schemes of data standardization are also implemented in the facility of {\em time series transformation}. Let the input time series from the current set be $\{x_t\}_{t=1}^N$ of length $N$ and the standardized time series be $\{y_t\}_{t=1}^N$. The four standardizations are defined as follows.
\begin{enumerate}
    \item Linear standardization: $y_t=\frac{x_t - x_{\mbox{\footnotesize{min}}}}{x_{\mbox{\footnotesize{max}}} - x_{\mbox{\footnotesize{min}}}}$. $\{y_t\}_{t=1}^N$ has minimum 0 and maximum 1.

    \item Normalized or $z$-score standardization: $y_t=\frac{x_t - \bar{x}}{s_x}$, where $\bar{x}$ is the mean and $s_x$ the standard deviation (SD) of $\{x_t\}_{t=1}^N$. $\{y_t\}_{t=1}^N$ has mean 0 and SD 1.

    \item Uniform standardization: $y_t=\hat{F}_x(x_t)$, where $\hat{F}_x$ is the sample marginal cumulative density function (cdf) of $\{x_t\}_{t=1}^N$ given by the rank order of each sample $x_t$ divided by $N$. The marginal distribution of $\{y_t\}_{t=1}^N$ is uniform in $[0,1]$.

    \item Gaussian standardization: $y_t=\Phi^{-1}\left (\hat{F}_x(x_t) \right)$, where $\Phi$ denotes the standard Gaussian (normal) cdf. The marginal distribution of $\{y_t\}_{t=1}^N$ is standard Gaussian.
\end{enumerate}
The Gaussian standardization can also serve as a transform to Gaussian time series.

\subsection{Resampling}
\label{subsec:Resampling}

When analyzing time series, it is often of interest to test an hypothesis for the underlying system to the given time series. The null hypotheses H$_0$ considered here are that a time series $\{x_t\}_{t=1}^N$ is generated by (a) an independent (white noise) process, (b) a Gaussian process and (c) a linear stochastic process. Many of the implemented measures in {\tt MATS} can be used as test statistics for these tests. The resampling approach can be used to carry out the test by generating randomized or bootstrap time series. The facility of {\em time series resampling} implements a number of algorithms for the generation of a given number of resampled time series for each H$_0$, as follows.

\begin{enumerate}
    \item {\em Random permutation} (RP) or shuffling of $\{x_t\}_{t=1}^N$ generates a time series that preserves the marginal distribution of the original time series but is otherwise independent. The RP time series is a randomized (or so-called surrogate) time series consistent to the H$_0$ of independence.

    \item It is conventionally referred to as {\em Fourier transform} (FT) algorithm because it starts with a Fourier transform of $\{x_t\}_{t=1}^N$, then the phases are randomized, and the resulting Fourier series is transformed back to the time domain to give the surrogate time series. The FT surrogate time series possesses the original power spectrum (slight inaccuracy may occur due to possible mismatch of data end-points). Note that the marginal distribution of the FT surrogate time series is always Gaussian and therefore it is consistent to the H$_0$ of a Gaussian process \citep{Theiler92a}.

    \item The algorithm of {\em amplitude adjusted Fourier transform} (AAFT) first orders a Gaussian white noise time series to match the rank order of $\{x_t\}_{t=1}^N$ and derives the FT surrogate of this time series. Then $\{x_t\}_{t=1}^N$ is reordered to match the ranks of the generated FT time series \citep{Theiler92a}. The AAFT time series possesses the original marginal distribution exactly and the original power spectrum (and thus the autocorrelation) approximately, but often the approximation can be very poor. Formally, AAFT is used to test the H$_0$ of a Gaussian process undergoing a monotonic static transform.

    \item The {\em iterated AAFT} (IAAFT) algorithm is an improvement of AAFT and approximates better the original power spectrum and may differ only slightly from the original one \citep{Schreiber96}. At cases, this difference may be significant because all generated IAAFT time series tend to give about the same power spectrum, i.e., the variance of the power spectrum is very small. The IAAFT surrogate time series is used to test the H$_0$ of a Gaussian process undergoing a static transform (not only monotonic).

    \item The algorithm of {\em statically transformed autoregressive process} (STAP) is used for the same H$_0$ as for IAAFT and the surrogate data are statically transformed realizations of a suitably designed autoregressive (AR) process \citep{Kugiumtzis02}. The STAP time series preserve exactly the original marginal distribution and approximately the original autocorrelation, so that the variance is larger than for IAAFT but there is no bias.

    \item The algorithm of {\em autoregressive model residual bootstrap} (AMRB) is a model-based (or residual-based) bootstrap approach that uses a fitted AR model and resamples with replacement from the model residuals to form the bootstrap time series \citep{Politis03,Hjellvik95}. AMRB is also used for the same H$_0$ as IAAFT and STAP. The AMRB time series have approximately the same marginal distribution and autocorrelation as those of the original time series, but deviations on both can be at cases significant.
\end{enumerate}

For a review on surrogate data testing, see \citet{Schreiber99b,Kugiumtzis01a}. For a comparison of the algorithms of AAFT, IAAFT, STAP and AMRB in the test for nonlinearity, see \citet{Kugiumtzis08a}.

The segmented, standardized and resampled time series get a specific coded name with a running index (following the name regarding the corresponding original time series) and they are stored in the current time series set. The name notation of the time series is given in Appendix A.

\subsection{Time series visualization}
\label{subsec:Viewtimeseries}

The facility of {\em time series visualization} provides one, two and three dimensional views of the time series, as well as histograms. Time history plots (1D plot) can be generated for a number of selected time series superimposing them all in one panel or plotting them separately but in one figure (ordering them vertically in one plot or in subplots). This can be particularly useful to see all segments of a time series together or surrogate time series and original time series together. Also, scatter plots in two and three dimensions (2D / 3D plots) can be generated for one time series at a time. This is useful when underlying deterministic dynamics are investigated and the user wants to get a projected view (in 2D and 3D) of the hypothesized underlying attractor. Histogram plots can be generated either as superimposed lines in one panel for the given time series, or as subplots in a matrix plot of given size, displaying one histogram per time series. In the latter case, the line of the fitted Gaussian distribution can be superimposed to visualize the fitting of the data to Gaussian and then the $p$-value of the Kolmogorov-Smirnov test for normality is shown as well.

\section{Measures on scalar time series}
\label{sec:Measures}

A large number of measures, from simple to sophisticated ones, can be selected. The measures are organized in three main groups: linear measures, nonlinear measures and ``other'' measures. In turn, each main group is divided in subgroups, e.g., the group of linear measures contains the subgroups of correlation measures, frequency-based measures and model-based measures. Many of the measures require one or more parameters to be specified and the default values are determined rather arbitrarily as different time series types (e.g., from discrete or continuous systems) require different parameter settings. For most free parameters the user can give multiple values or a range of values and then the measure is computed for each of the given parameter values.

The user is expected to navigate over the different measure groups, select the measures and the parameter values and then start the execution, i.e., the calculation of each selected measure (possibly repeated for different parameter values) on each time series in the current time series set. Execution time can vary depending on the number and length of the time series as well as the computational requirements of the selected measures. Some nonlinear measures, such as the correlation dimension and the local linear fit and prediction are computationally rather involved and may require long execution time. When execution is finished the output is stored  in a matrix and the name of each calculated measure for the specific set of free parameters is contained in the {\em current measure list}. The name notation for measures and parameters is given in Appendix B. Each such name regards a row in the output matrix, i.e., an array of values of length equal to the number of time series for which the measure was executed. The names of these time series constitute the {\em measured time series list}, which regards the current time series set at the moment of execution. Note that the current time series set may change after the execution of the measures but the measured time series list is preserved for reference purposes until the measures are executed again. In the following, the measures are briefly presented.

\subsection{Linear measures group}
\label{subsec:Linear}
The group of linear measures includes 17 different measures regarding linear correlation, frequency and linear autoregressive models.

The correlation subgroup includes measures of linear and monotonic autocorrelation, i.e., in addition to measures based on the \emph{Pearson autocorrelation} function measures based on the \emph{Kendall} and \emph{Spearman autocorrelation} functions are included as well \citep{Hallin92}.
Note that the autocorrelation of any type is considered as a different measure for each delay $t$, so that for a range of delay values equally many measures of the selected type are generated.
In addition to autocorrelation, we introduce the measures of
{\em cumulative autocorrelation} defined by the sum of the magnitude of the autocorrelations up to a given maximum lag. Moreover, specific lags that correspond to zero or $1/e$ autocorrelation are given as additional correlation measures. In the same group, the partial autocorrelation is included for delay $t$ because it regards the linear correlation structure of the time series. The partial autocorrelation is used to define the order of an autoregressive (AR) model by the largest lag that provides significant partial autocorrelation (the user can use the visualization facility of measure vs parameter to assess this).

The subgroup of frequency-based measures consists of the {\em energy fraction in bands} and the {\em median frequency}, computed from the standard periodogram estimate of the power spectrum. The energy fraction in a frequency band is simply the fraction of the sum of the values of the power spectrum at the discrete frequencies within the band to the energy over the entire frequency band. Up to 5 energy fractions in bands can be calculated. The median frequency is the frequency $f^*$, for which the energy fraction in the band $[0,f^*]$ is half. These frequency measures have been used particularly in the analysis of EEG \citep{Gevins87}.

The standard linear models for time series are the {\em autoregressive  model} (AR) and {\em autoregressive moving average model} (ARMA) \citep{Box94a}. Statistical measures of fitting and prediction with AR or ARMA can be computed for varying model orders (for the AR part and the moving average, MA, part) and prediction steps. The difference between the fitting and prediction measures lies in the set of samples that are used to compute the statistical measure: all the samples are used for fitting and the samples in the so-called test set are used for prediction. Here, the test set is the last part of the time series. For multi-step ahead prediction (respectively for fit) the iterative scheme is used and the one step ahead predictions from previous steps are used to make prediction at the current step. Four standard statistical measures are encountered: \emph{the mean absolute percentage error} (MAPE), the \emph{normalized mean square error} (NMSE), the \emph{normalized root mean square error} (NRMSE) and the \emph{correlation coefficient} (CC). All four measures account for the variance of the time series and this allows comparison of the measure across different time series.

\subsection{Nonlinear measures group}
\label{subsec:Nonlinear}
The group of nonlinear measures includes 19 measures based on the nonlinear correlation (bicorrelation and mutual information), dimension and complexity (embedding dimension, correlation dimension, entropy) and nonlinear models (local average and local linear models). A brief description of the measures in each class follows below.

\subsubsection{Nonlinear correlation measures}
\label{subsec:Noncor}

Two measures of nonlinear correlation are included, the bicorrelation and the mutual information. The {\em bicorrelation}, or three-point autocorrelation, or higher order correlation, is the joint moment of three variables formed from the time series in terms of two delays $t$ and $t^{\prime}$. A simplified scenario for the delays is implemented, $t^{\prime} = 2t$, so that the measure becomes a function of a single delay $t$. The bicorrelation is then defined by $E[x(i), x(i+t), x(i+2t)]$, where the mean value is estimated by the sample average (the quantity is standardized in the same way as for the second order correlation, i.e., Pearson autocorrelation).

The {\em mutual information} is defined for two variables $X$ and $Y$ as the amount of information that is known for the one variable when the other is given, and it is computed from the entropies of the variable vector $[X,Y]$ and the scalar variables $X$ and $Y$. For time series, $X = x(i)$ and  $Y = x(i+t)$ for a delay $t$, so that the mutual information is a function of the delay $t$. Mutual information can be considered as a correlation measure for time series that measures the linear and nonlinear autocorrelation. There are a number of estimates of mutual information based on histograms, kernels, nearest neighbors and splines. There are advantages and disadvantages of all estimates, and we implement here the histogram-based estimates of {\em equidistant} and {\em equiprobable binning} because they are widely used in the literature \citep[e.g., see ][]{Cellucci05,Papana08}.

Similarly to the correlation measures, the measures of {\em cumulative bicorrelation, cumulative equidistant mutual information} and {\em cumulative equiprobable mutual information} at given maximum delays are implemented as well. The {\em first minimum of the mutual information} function is of special interest and the respective delay can be used either as a discriminating measure or as the optimal delay for state space reconstruction, a prerequisite of nonlinear analysis of time series. This specific delay is computed for both the equidistant and the equiprobable estimate of mutual information.

The bicorrelation is not a widely discussed measure, but the cumulative bicorrelation has been used in the form of the statistic of the the so-called Hinich test of linearity (or non-linearity as it is best known in the dynamical systems approach of time series analysis), under the constraint that the second delay is twice the first delay \citep{Hinich96}. The bicorrelation, under the name three point autocorrelation, has been used as a simple nonlinear measure in the surrogate data test for nonlinearity \citep{Schreiber97,Kugiumtzis01}.

\subsubsection{Dimension and complexity measures}
\label{subsec:Nondimcomp}

Nonlinear dynamical systems are characterized by invariant measures, which can be estimated from observed time series. The invariant measures are associated with the complexity of the underlying system dynamics to time series and the dimension of the system's attractor, i.e., the set of points (trajectory) generated by the dynamical system.

The \emph{correlation dimension} is a measure of the attractor's fractal dimension and it is the most popular among other fractal dimension measures, such as the information and box-counting dimension \citep{Grassberger83}. The correlation dimension for a regular attractor, such as a set of finite points (regarding a periodic trajectory of a map), a limit cycle and a torus, is an integer but for a so-called strange attractor possessing a self-similar structure is a non-integer. The correlation dimension is estimated from the sample density of the distances of points reconstructed from the time series, typically from delay embedding for a delay $t$ and an embedding dimension $m$. The estimation starts with the computation of the so-called correlation sum $C(r)$ (abusively, it is also called correlation integral), which is calculated for a range of distances $r$. The latter is here referred to as radius because the Euclidean norm is used to compute the inter-point distances of all points from each reference point. Thus for each $r$, $C(r)$ is the average fraction of points within this radius $r$. Given that the underlying dynamical system is deterministic, the correlation sum scales with $r$ as $C(r)$ $\sim r^{\nu}$, and the exponent $\nu$ is the correlation dimension. Theoretically, the scaling should hold for very small $r$, but in practice the scaling is bounded from below by the lack of nearby points (dependent on the length of the time series) and the masking of noise (even for noise-free data a lower bound for $r$ is set by the round-off error).

Computationally, for the estimation of $\nu$, first the local slope of the graph of $\log(C(r))$ versus $\log(r)$ is computed by the smooth derivative of $\log(C(r))$ with respect to $\log(r)$. Provided that the scaling holds for a sufficiently large range of (small) $r$, a horizontal plateau of the local slope vs $\log(r)$ is formed for this range of $r$. The least size of the range is usually given in terms of the ratio of the upper edge $r_2$ to the lower edge $r_1$ of the scaling $r$-interval, and a commonly used ratio is $r_2/r_1 = 4$. Here, the estimate of $\nu$ is the mean local slope in the interval $[r_1,r_2]$ found to have the least variance of the local slope.

In practice, reliable estimation of $\nu$ is difficult since there is no clear scaling, meaning that the variance of the local slope is large. In these cases, other simpler measures derived from the correlation sum may be more useful, such as the value of the correlation sum $C(r)$ for a given $r$, which is simply the \emph{cumulative density} function of inter-point distances at a given inter-point distance, or the value of $r$ that corresponds to a given $C(r)$, which is the \emph{inverse of this cumulative density} \citep{McSharry03,Andrzejak06}.

The embedding dimension $m$ is an important parameter for the analysis of time series from the dynamical system approach, and determines the dimension of the Euclidean pseudo-space in which (supposedly) the attractor is reconstructed from the time series. An $m$ that is small but sufficiently large to unfold the attractor is sought and a popular method to estimate it is the method of \emph{false nearest neighbors} (FNN) \citep{Kennel92}. The method evaluates a criterion of sufficiency for increasing $m$. For each reconstructed point at the examined embedding dimension $m$, it is checked whether its nearest neighbor falls apart with the addition of a new component, i.e., the distance increase at least by a so-called escape factor when changing the embedding dimension from $m$ to $m+1$. If this is so, the two points are called false nearest neighbors. The estimated minimum embedding dimension is the one that first gives an insignificant percentage of false nearest neighbors. In our FNN algorithm we use the Euclidean metric of distances and two fixed parameters in order to deal with the problem of occurrence of a large distance of the target point to its nearest neighbor \citep{Hegger99}. The first parameter sets a limit to this distance and the second sets a limit to the lowest percentage of target points for which a nearest neighbor could be found, in order to maintain sufficient statistics. As a result the computation may fail for large embedding dimensions relative to the time series length (then the output measure value is blank). We believe that this is preferable than obtaining unreliable numerical values that can mislead the user to wrong conclusions. Though the minimum $m$ estimated by this method is not an invariant measure, as it depends for example on the delay parameter $t$, it is used in some applications as a measure to discriminate different dynamical regimes and it is therefore included here as well.

Among different complexity measures proposed in the literature, the \emph{approximate entropy} and the \emph{algorithmic complexity} are implemented here. The approximate entropy is a computationally simple and efficient estimate of entropy that measures the so-called ``pattern similarity'' in the time series \citep{Pincus91}. It is expressed as the logarithm of the ratio of two correlation sums (for an appropriately chosen radius $r$) computed for embedding dimensions $m$ and $m+1$.

Formally, the algorithmic complexity (also referred to as Kolmogorov complexity) quantifies how complex a symbolic sequence is in terms of the length of the shortest computer program (or the set of algorithms) that is needed to completely describe the symbolic sequence. Here, we follow the approach of Lempel and Ziv who suggested a measure of the regularity of the symbolic sequence, or its resemblance to a random symbolic sequence \citep{Lempel76}. Computationally, the complexity of the sequence is evaluated by the plurality of distinct words of varying length in the sequence of symbols.
The algorithmic complexity measure, though defined for symbolic sequences, has been used in time series analysis, e.g., see \citet{Radhakrishnan00, Zhao06}. Similarly to the mutual information, the time series is discretized to a number of symbols assigned to the allocated bins, which are formed by equidistant and equiprobable binning.

There are a number of other measures of complexity, most notably the largest Lyapunov exponent, that are not implemented here. In our view the existing methods for the estimation of the largest Lyapunov exponent do not provide stable estimates as the establishment of scaling of the exponential growth of the initial perturbation requires large and noise-free time series, e.g., see the methods implemented in {\tt TISEAN} \citep{Hegger99b}.

\subsubsection{Nonlinear model measures}
\label{subsec:Nonmodel}

Among different classes of nonlinear models, the local state space models are implemented here\footnote{For example, {\tt MATLAB} provides a toolbox for neural networks, a large and popular class of linear and nonlinear models.}. This class is popular in nonlinear analysis of time series because it is effective for fitting and prediction tasks and it is intuitively simple \citep{Abarbanel96, Kantz97}.
A model is formed for each target point $\mathbf{x}_i$ reconstructed from the time series with a delay $t$ and an embedding dimension $m$. The local region supporting the model is determined here by the $k$ nearest neighbors of $\mathbf{x}_i$, denoted as $\mathbf{x}_{i(j)}$, for $j=1,...,k$, found in the available set of reconstructed points (different for fitting and prediction). The $k$-$D$ tree data structure is utilized to speed up computation time in the search of neighboring points, as implemented by the contributed {\tt MATLAB} software \citep{Shechter04}.

Three different types of local models are implemented and they are specified by the value of the so-called truncation parameter $q$, as follows. Assuming a current state $\mathbf{x}_i$ the task is to make $h$-step ahead predictions.
\begin{enumerate}
    \item The {\em local average model} (LAM), called also model of zeroth order, predicts $x_{i+h}$  from the average of the mappings of the neighbors at lead time $h$, i.e., $x_{i(j)+h}$, for $j=1,...,k$. This model is selected by setting $q = 0$.
    \item The {\em local linear model} (LLM) is the linear autoregressive model $x_{i+h} = a_0 + \mathbf{a}^T \mathbf{x}_i$ that holds only for a region around the current point $\mathbf{x}_i$ formed by the neighboring points $\mathbf{x}_{i(j)}$, for $j=1,...,k$ \citep{Farmer87}. The parameters $a_0$ and $\mathbf{a}$ are estimated by \emph{ordinary least squares} (OLS) and this requires that $k \ge m+1$. The prediction $x_{i+h}$ is computed from the evaluation of the model equation for the target point $\mathbf{x}_i$. Note that the solution for the model parameters may be numerically unstable if $k$ is close to $m$. This model is selected by using any $q \ge m$.
    \item The OLS solution for the parameters of the local linear model is regularized using the {\em principal component regression} (PCR) \citep{Kugiumtzis02b,Kugiumtzis98}. PCR rotates the natural basis of local state space to match the basis formed by the principal components found by Singular Value Decomposition (SVD) of the matrix formed by the neighboring points. Then the space is projected to the subspace formed by the $q$ first principal axes, the solution for the parameters is found in this subspace and it is transformed back to the original state space to yield the PCR regularized solution for the parameters. In this way, the estimated parameters have smaller variance (they are more stable) at the cost of introduced bias. Another advantage is that PCR may reduce the effect of noise. The condition for the PCR solution is that $k > q+1$, so that stable solutions can be reached even when $m > k$ provided that the truncation parameter is sufficiently small. This solution is selected if $0 < q < m$.
\end{enumerate}

The fitting and prediction measures from local models are the same as for the linear models, namely MAPE, NMSE, NRMSE and CC. For multi-step fit or prediction, both the direct and iterative schemes are implemented. For the direct fit or prediction at a lead time $h$, the $h$-step ahead mappings are used to form the estimate of the target points at a lead time $h$, whereas for the iterative fit or prediction, the one step predictions from previous steps are used to make predictions for the current step until the $h$-step is reached.

\subsection{Other measures}
\label{subsec:other}

This last group includes 16 measures divided into two subgroups: moment and long range correlation measures, and feature statistics.

The first subgroup does not represent a distinct class but it is rather a summary of descriptive statistics of the time series ({\em mean, median, variance, standard deviation, interquartile distance, skewness and kurtosis}), measures of \emph{moments of the first and second differences} of the time series (the Hjorth parameters) and measures of long range correlation (\emph{rescaled range analysis}, R/S, and \emph{detrended fluctuation analysis}, DFA). The \emph{Hjorth parameters of mobility and complexity} are simple measures of signal complexity based on the second moment of the signal and its first and second difference. These measures have been used in the electroencephalography (EEG) analysis and they are ``... clinically useful tools for the quantitative description of an EEG'', as quoted from the original paper in \citet{Hjorth70}.  On the other hand, R/S analysis and DFA are methods for the estimation of the so-called Hurst exponent, an index of long range correlation in time series, and they have been used mostly in economics and geophysics \citep{Edgar96, Mandelbrot97,Peng94}. Both R/S analysis and DFA estimate the scaling of the variance (or energy) of the time series at different time scales, with DFA being robust to the presence of trends in the time series (we implement the DFA of order one, meaning that linear polynomials are fitted to remove trends).

The measures of the feature statistics subgroup regard descriptive statistics of oscillation characteristics and are meaningful for oscillating time series. Each oscillation can be roughly, but often sufficiently, described by four features, the peak (\emph{local maximum}) and trough (\emph{local minimum}) of the oscillation, the \emph{time duration of the oscillation} (e.g., from one peak to the next) and the \emph{time from peak to trough} (or equivalently from trough to peak). A fifth feature that can be useful at cases, especially in the presence of drifts in the time series, is formed by the \emph{difference of a peak and the corresponding trough}. The first two features constitute the turning points of the time series and have been used in the analysis of oscillating time series \citep[e.g., see ][]{Kugiumtzis04a} and even non-oscillating time series, e.g., in finance \citep{Garcia98,Bao08}.
For the computation of the features the turning points have to be identified first. A sample point of the time series is a turning point, i.e., local minimum (maximum), if it is the smallest (largest) of the samples of the time series in a data window centered at this sample point. The data window size is given as $2w+1$ for a given offset value $w$. For noisy time series, smoothing prior to the turning point detection is suggested and this is implemented here with a moving average filter of a given order in forward and backward direction. Standard descriptive statistics are given as measures from each of the five oscillating features. The use of statistics of oscillating features as measures of time series analysis is rather new and it has been recently introduced in \citet{Kugiumtzis06, Kugiumtzis07}.

\section[The structure of MATS]{The structure of {\tt MATS}}
\label{sec:software}

\begin{figure}[htb] 
\centerline{\includegraphics[width=12cm,keepaspectratio]{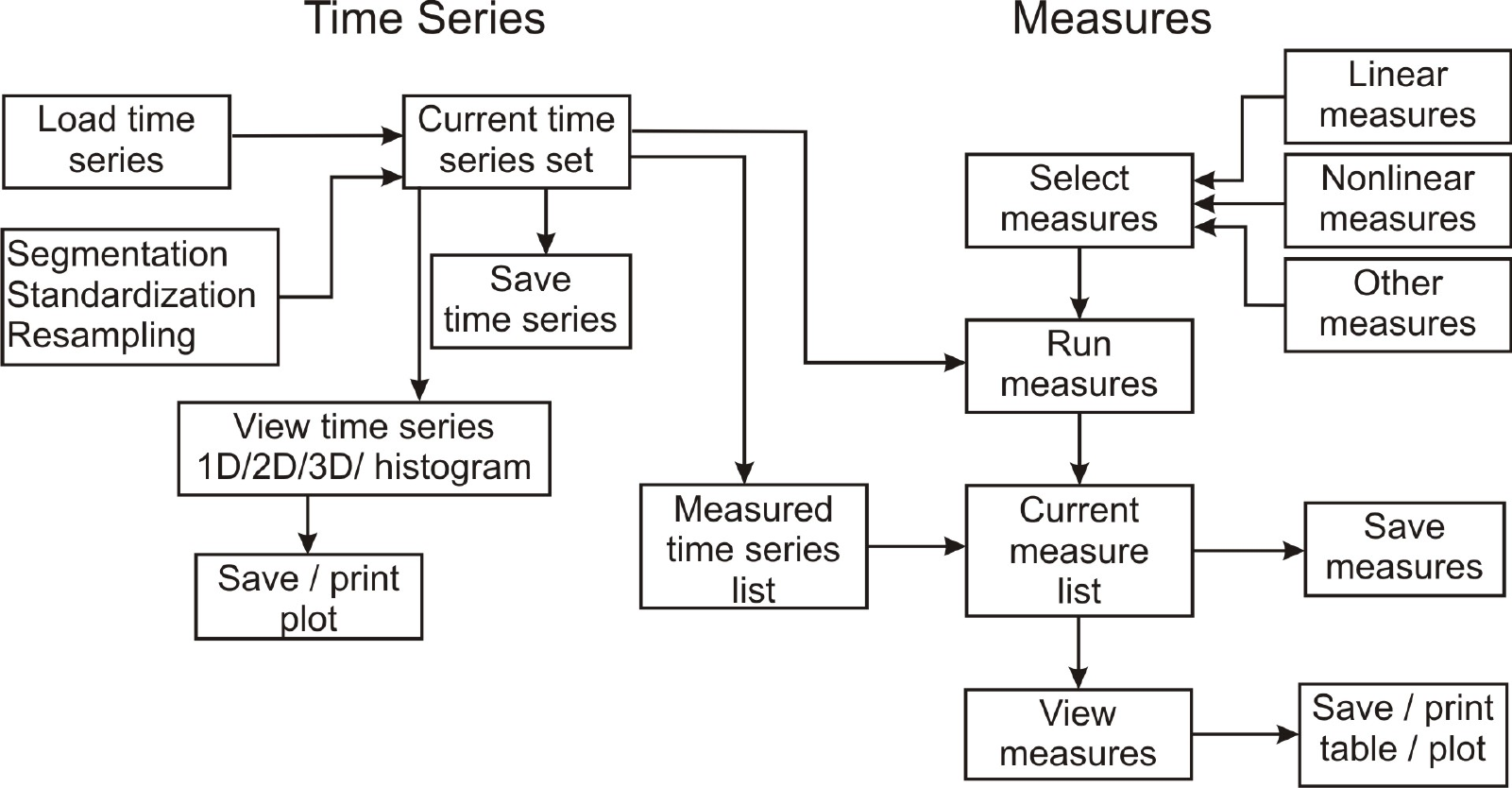}}
 \caption{Flow diagram for the possible operations in {\tt MATS}.}
\label{flowdg}
\end{figure}

\begin{figure}[htb] 
\centerline{\includegraphics[width=16cm,keepaspectratio]{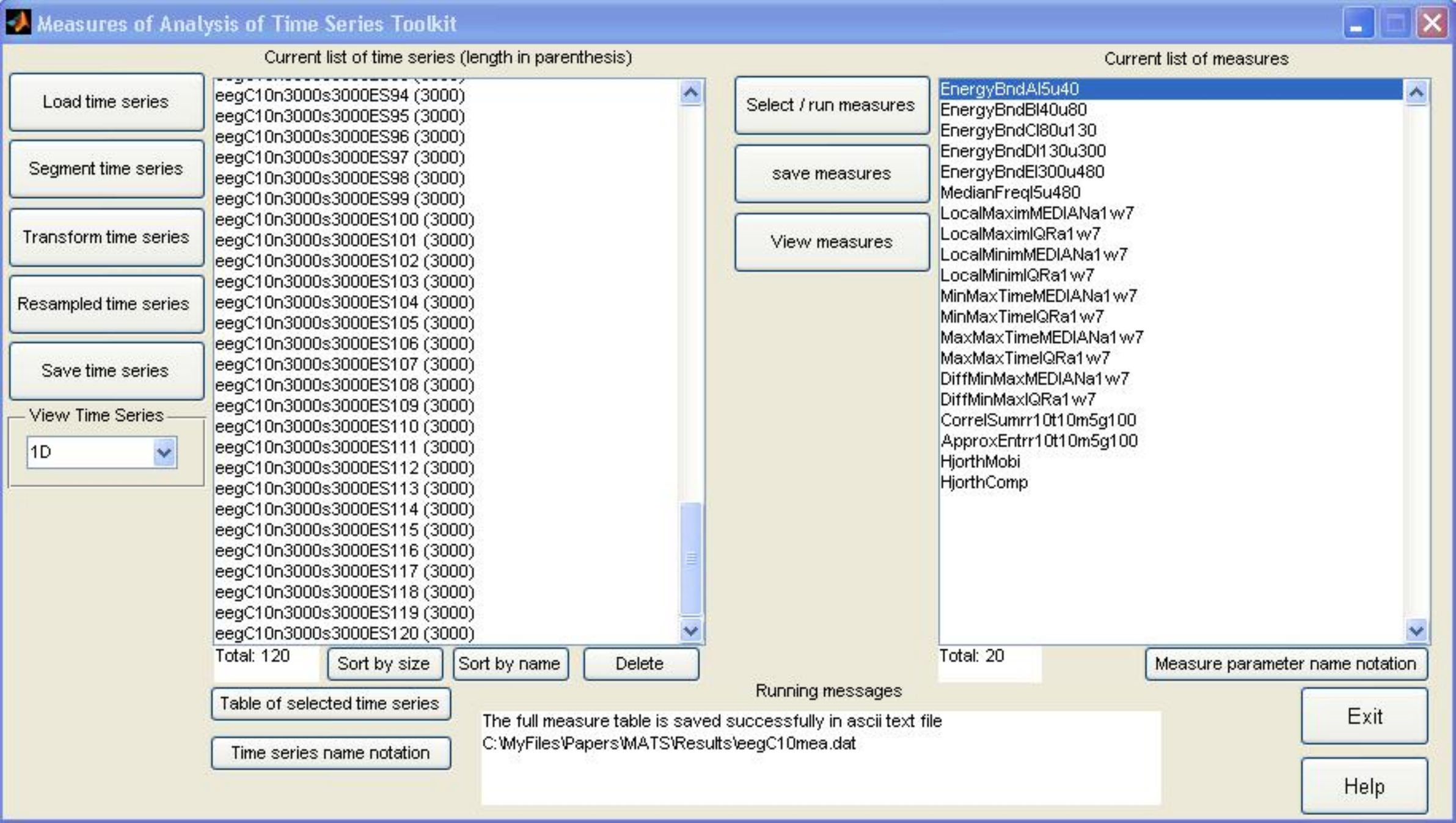}}
 \caption{Screen shot of the main menu of {\tt MATS}. The shot was taken after the measure calculations for the example 1 in Section~\ref{sec:Applications} were completed and the measure table was saved, as denoted in the running message box at the bottom of the window.}
\label{mainmenu}
\end{figure}

A brief overview of the structure and the main features of the {\tt MATS} toolkit is given below. The user interface consists of several components that are integrated in two modules, one for time series handling and one for measure calculation. For each module a data list is assigned, i.e., the list of the current time series set (referred to as the current time series list) and the current measure list, respectively.
Figure~\ref{flowdg} shows the flow diagram of the basic operations of {\tt MATS} in each of the two modules and additional features for importing, exporting and presentation (tables and figures) of time series and measures.
Moreover, in Figure~\ref{mainmenu} a screen-shot of {\tt MATS} main window is shown, where the selections and the list for each of the two modules are organized on the left and right of the window.

\subsection{Time series handling}

All the GUI components of the first module are designed in the same way and they access and update a commonly shared data base, which is communicated to the end user through the current time series list. This data base is globally declared in the main GUI and is directly and effectively accessed by the GUI components (for this the {\tt MATLAB} commands {\tt setappdata} and {\tt getappdata} are used).

The current time series list is dynamic; it is formed and changed using the operations for loading, segmenting, transforming and resampling time series. The user can also delete time series from the list and sort the displayed list with respect to name and size.

\subsection{Measure selection and calculation}
\label{subsec:MeaSelection}

The GUI components for the module of measure selection are organized in groups and subgroups. At each GUI window the user is opted to select the corresponding measures and their parameters.
A measure is selected by check box activation and then the measure parameters (if any) become active and their default values can be changed.
The configuration of the current selected measures and parameters can be stored in a file for later use, meaning that files of configurations of selected measures in special format can also be imported.

When the user completes the selection of measures, the user may start the execution, i.e., the computation of the selected measures on all the time series in the current time series list. At this point, the names in the current time series list are passed to the measured time series list. Further, the user may alter the current time series list, but the measured time series list is intact and can be called in order to assign the measure values to the corresponding time series for visualization or exporting purposes.

Many of the considered measures in {\tt MATS} are functions of one or more free parameters. It is stressed that the term {\em measure} is used in {\tt MATS} to denote the specific outcome of the function for given values of the free parameters. So, the selection of one function can actually give rise to a number of measures (as large as the number of the given parameter values). Each such a measure is given a unique name consisting of the measure code name and each parameter code name followed by the corresponding parameter value. The names of the measures in the current measure list are displayed in the main window of {\tt MATS} (see Figure~\ref{mainmenu}). The measure values for the different time series constitute the second data base that is accessed from the GUI components in the same way as for the data base for the time series.

\begin{figure}[htb] 
\centerline{\includegraphics[width=16cm,keepaspectratio]{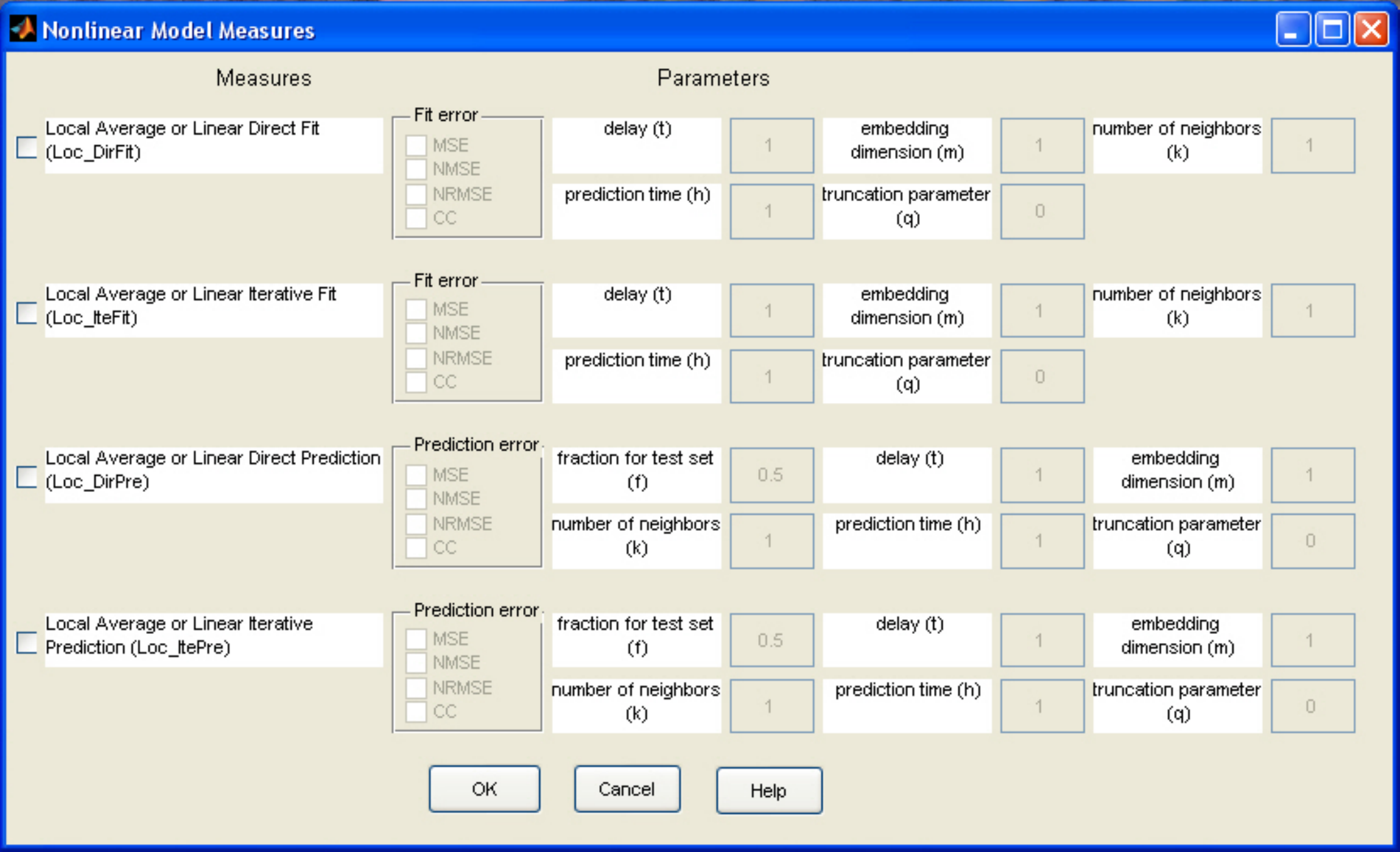}}
 \caption{Screen-shot of the GUI named {\tt Nonlinear Model Measures} called in the GUI {\tt Measure Selection}.}
\label{meagroup}
\end{figure}

An example of measure selection is given below for the nonlinear model measures and the corresponding screen-shot is shown in Figure~\ref{meagroup}.
The user can select one or more of the local fit and prediction measures (making use of the direct or iterative scheme). Suppose the user activates the check box of the first model
(``Local Average or Linear Direct Fit'' with 10 character code name {\tt Loc\_DirFit}). Then the fields for the parameters are highlighted and the user can specify one or more of the four fit error statistics, or change the default parameter values. Some parameters bear only a single value, such as {\tt q} for specifying the type of local model, whereas other parameters can take a range of valid values. The validity of the given values in each field is checked and appropriate error messages are displayed if wrong {\tt MATLAB} syntax is used or the values are out of the valid range. For example, for the embedding dimension {\tt m} and the prediction time {\tt h}, valid values are any positive integers. So if we give in the field for {\tt m}, {\tt 1 3 5:5:20}, for {\tt h}, {\tt 1:5}, set all the other parameters to single values and select one fit error statistic, then upon execution of the selected measures, $30$ measures of {\tt Loc\_DirFit} will be added in the current measure list.

\subsection{Measure visualization}
\label{subsec:ViewMea}

An elaborate facility of {\tt MATS} is the set of functions for visualizing the computed measures on the ensemble of time series. The user can make use of standard {\tt MATLAB} functions to edit, print, and/or save the resulting tables and plots in a variety of formats.

\begin{figure}[htb] 
\centerline{\includegraphics[width=5cm,keepaspectratio]{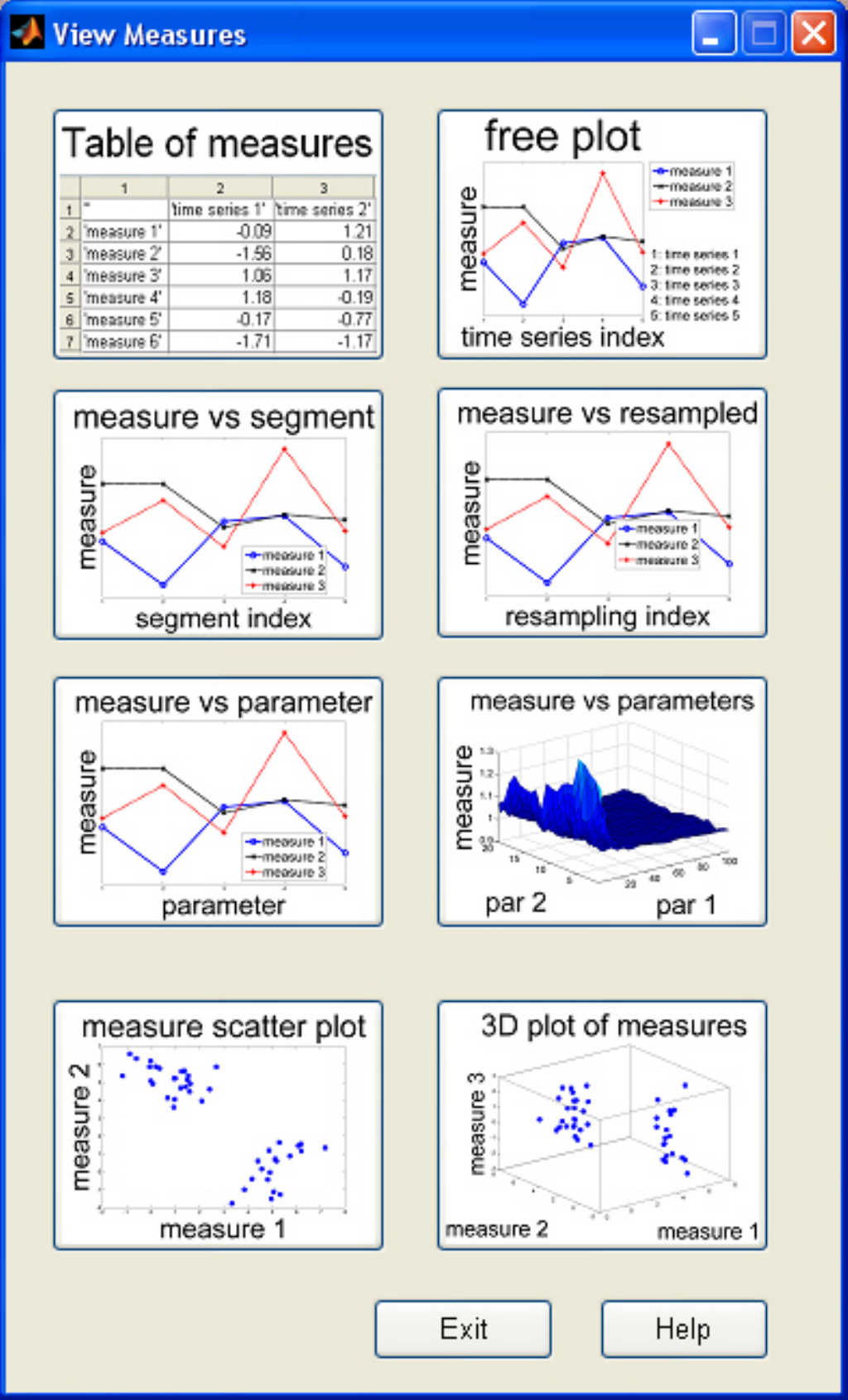}}
 \caption{Screen-shot of the GUI {\tt View Measures} that opts for seven different plot types and one table.}
\label{meaviewfig}
\end{figure}

The different visualizations are easily specified and executed as the user selects measure names from the current measure list and time series names from the measured time series list. The different visualization facilities are shown in the screen-shot of Figure~\ref{meaviewfig}.

The whole content of the current measure list or parts of it can be viewed in a table or in the so-called ``free-plot''. When measures are computed on segments of time series the plot of measures vs segment indices can be selected. The displayed list of time series in the interface window for plotting is a subset of the measured time series list and includes only names of time series generated by the segmenting facility (having names containing indices after the character {\tt S}). The GUI for the plot of measures vs resampled time series is designed similarly. For the other plot facilities, all the names of the measured time series list are displayed and can be selected. In particular, in the displayed list of time series for plotting measure vs resampled time series, also the original time series is included and is given the index $0$ in the plot. In addition, a parametric test (assuming normal distribution of the measure values on the resampled data) and a nonparametric test (using rank ordering of the measure values on the original and surrogate data) are performed for each measure and the $p$-values are shown in the plots \citep[for details, see][]{Kugiumtzis01a}. To assess the validity of the parametric test, the $p$-value of the Kolmogorov-Smirnov test for normality is shown as well.

The GUI that makes a plot of measure vs parameter (2D plot) allows the user to select a parameter from the list of all parameters and mark the measures of interest from the current measure list. Even if there are irrelevant measures within the selected ones, i.e., measures that do not include the key character for the selected parameter, they will be ignored after checking for matching the parameter character in the measure name. The same applies for two parameters in the GUI for the 3D plot of measure vs two parameters. Finally, the GUI of measure scatter plot makes a 2D scatter plot of points (each point regards a time series) for two selected measures and another GUI makes a 3D scatter plot for three selected measures.

All different plots can be useful for particular types of time series analysis and some representative applications of {\tt MATS} are given in the next Section.

\section[Application of MATS]{Application of {\tt MATS}}
\label{sec:Applications}

{\tt MATS} toolkit can be used in various applications of time series analysis such as the following.

\begin{itemize}

\item Detection of regime change in long data records: allowing for computing a measure on consecutive segments of a long data record and detecting on the series of measure values abrupt or smooth changes as well as trends.

\item Surrogate data test for nonlinearity: allowing for the selection of different surrogate data generating algorithms and many different test statistics.

\item Discrimination ability of different measures: comparing different measures with respect to their power in discriminating different types of time series.

\item Assessing the dependence of a measure on measure specific parameters: providing graphs of a measure versus its parameters and comparing the dependence of the measure on its parameters for different types of time series.

\item Feature-based clustering of time series data base: computing a set of measures (features) on a set of time series (for two or three features the clusters can be seen in a 2D or 3D plot). For quantitative results, the array of measure data has to be fed in a feature-based clustering algorithm (we currently work on developing such a tool and link it to {\tt MATS}).

\end{itemize}

Two examples are given below to illustrate the use of {\tt MATS}. The data in the examples are from epileptic human EEG but it is obvious that {\tt MATS} can be applied on any simulated or real world time series.

\paragraph{Example 1}

In the analysis of long EEG records the main interest is in detecting changes in the EEG signals that may presignify an impending event, such as seizure. This analysis is easily completed in {\tt MATS} and consists of data segmentation, measure selection and computation, and finally visualization. The record in this example covers one hour of human EEG from $63$ channels (system $10-10$) sampled at $100$ Hz and contains a seizure at $47$ min and $40$ sec. The name of the file is {\tt eeg.dat} (in ascii format). The first $10$ channels will be visualized and the tenth channel will be further processed for illustration of the measure profiles on consecutive segments from this channel.

\begin{figure}[htb] 
\centerline{\includegraphics[width=10cm,keepaspectratio]{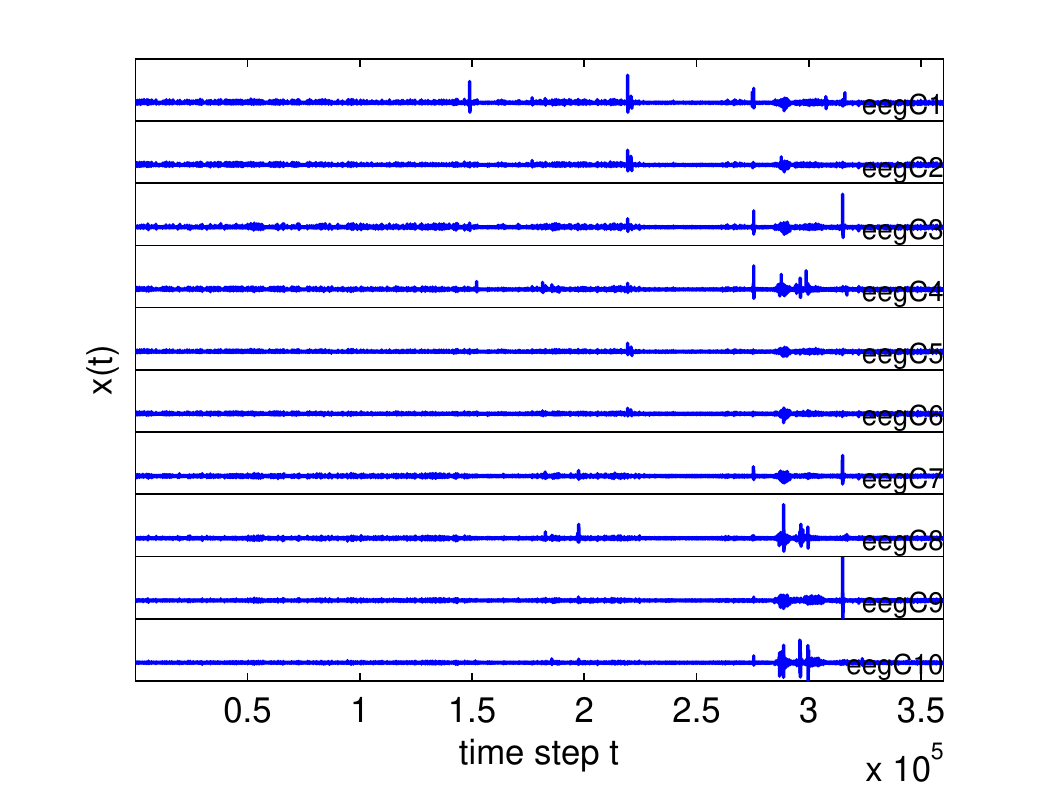}}
 \caption{A plot of EEG from the first $10$ channels of an hour long record sampled at $100$ Hz as generated by {\tt View Time Series - 1D}. Note that the seizure onset is at about $48$ min, i.e., at time step $288000$.}
\label{fig:eegser}
\end{figure}

{\em Step 1. Data segmentation}: The first $10$ channels of the record {\tt eeg.dat} are loaded and passed to the current time series list. Then the displayed list in the main menu contains ten names, from {\tt eegC1} to {\tt eegC10}. A plot of the ten time series using the facility {\tt View Time Series} is shown in Figure~\ref{fig:eegser}.

\begin{figure}[htb] 
\centerline{\includegraphics[width=10cm,keepaspectratio]{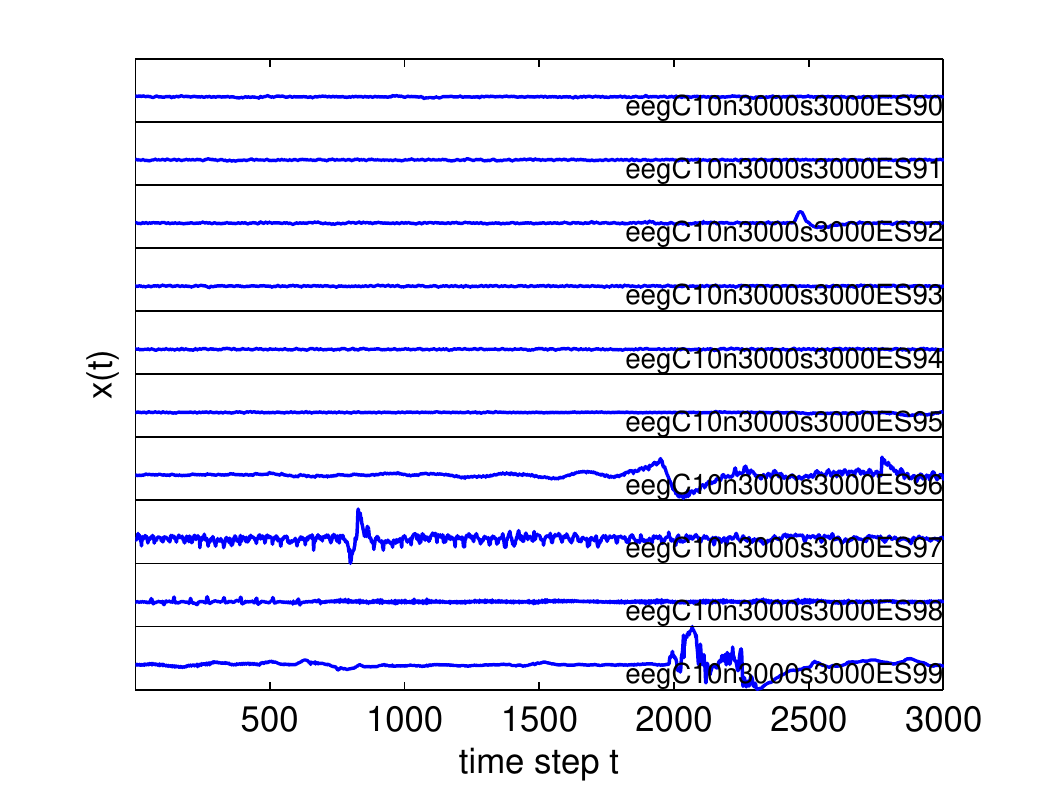}}
 \caption{A plot of the segmented time series with running index from $90$ to $99$, as generated by {\tt View Time Series - 1D}. Note that the seizure onset is about the middle of segment $96$.}
\label{fig:preictalC10seg}
\end{figure}

In all channels the seizure onset can be seen by a change in the amplitude of fluctuations of the EEG. The objective here is to investigate if there is any progressive change in the EEG at the pre-ictal period and this cannot be observed by eyeball judgement in any of the $10$ channels in Figure~\ref{fig:eegser}. We intend to investigate whether a measure profile can indicate such a change. For this we segment the data and compute measures at each segment. We choose to proceed with the $10$th channel, and make consecutive non-overlapping segments of $30$ sec. Using the segmentation facility and setting $3000$ for the length of segment (denoted by the symbol {\tt n}) and the same for the sliding step (denoted by the symbol {\tt s}), $120$ new time series are generated with the names {\tt eegC10n3000s3000ES1} to {\tt eegC10n3000s3000ES120}, where the character {\tt E} stands for the option of ignoring remaining data from the end. Finally, we delete the $10$ long time series and the current time series list contains the $120$ segmented time series. A plot of the time series indexed from $90$ to $99$ is shown in Figure~\ref{fig:preictalC10seg} using again the {\tt View Time Series} facility. Note that the seizure starts at segment indexed by $96$.

{\em Step 2. Measure selection and computation}: Using the facility {\tt Select / run measures} and with appropriate selections the following measures are activated to be computed on the $120$ segmented time series: energy bands, correlation sum, approximate entropy, Hjorth parameters and all feature statistics. For the correlation sum and approximate entropy, the parameters were set as follows: radius $\mbox{{\tt r}}=0.1$, delay $\mbox{{\tt t}}=10$, embedding dimension $\mbox{{\tt m}}=5$ and Theiler window $\mbox{{\tt g}}=100$. For the energy bands, the frequency intervals of the power spectrum were selected to correspond to the $\delta$, $\theta$, $\alpha$, $\beta$ and $\gamma$ waves and the median frequency was computed as well (for a range of normalized frequencies from $0.005$ to $0.48$). These waves have been extensively studied in encephalography research and are found to characterize various brain activities like sleep stages, but their role in pre-ictal activity is not established \citep{Gevins87}. Hjorth parameters are easily computed measures that have also been used in brain studies for very long \citep{Hjorth70}. The correlation sum and approximate entropy are used also in EEG analysis under the perspective of nonlinear dynamical systems \citep{Pincus91,McSharry03,Andrzejak06}.
The statistics of oscillating features are recently introduced as simple alternatives for capturing changes in the oscillation patterns of the EEG signal \citep{Kugiumtzis06, Kugiumtzis07}. For the detection of turning points, the parameter of offset for the local window ({\tt w}) was set to $7$ and the moving average filter order ({\tt a}) was set to $1$. The statistics median and inter-quartile range (IQR) were selected for all five oscillating features.

Upon completion of the computation of the measures on all $120$ time series in the current time series list, the current measure list is filled as shown in the main menu window in Figure~\ref{mainmenu}. It can then be saved for later use and this is actually done just before the snapshot of Figure~\ref{mainmenu} was taken, as denoted in the field titled ``Running Messages''. Note that the parameter settings are included in the names of the measures to avoid any confusion with regard to the application setup of the measures. For example, the first energy band at the top of the current measure list (see Figure~\ref{mainmenu}) is denoted as {\tt EnergyBndAl10u40}, where the first $10$ characters denote the measure name (the last character {\tt A} stands for the first band), {\tt l10} denotes that the lower value of the frequency band is $10/1000=0.01$, i.e., $1$ Hz, and {\tt u40} denotes that the upper value of the frequency band is $40/1000=0.04$, i.e., $4$ Hz.

\begin{figure}[htb] 
\centerline{\hbox{
  \includegraphics[width=7cm,keepaspectratio]{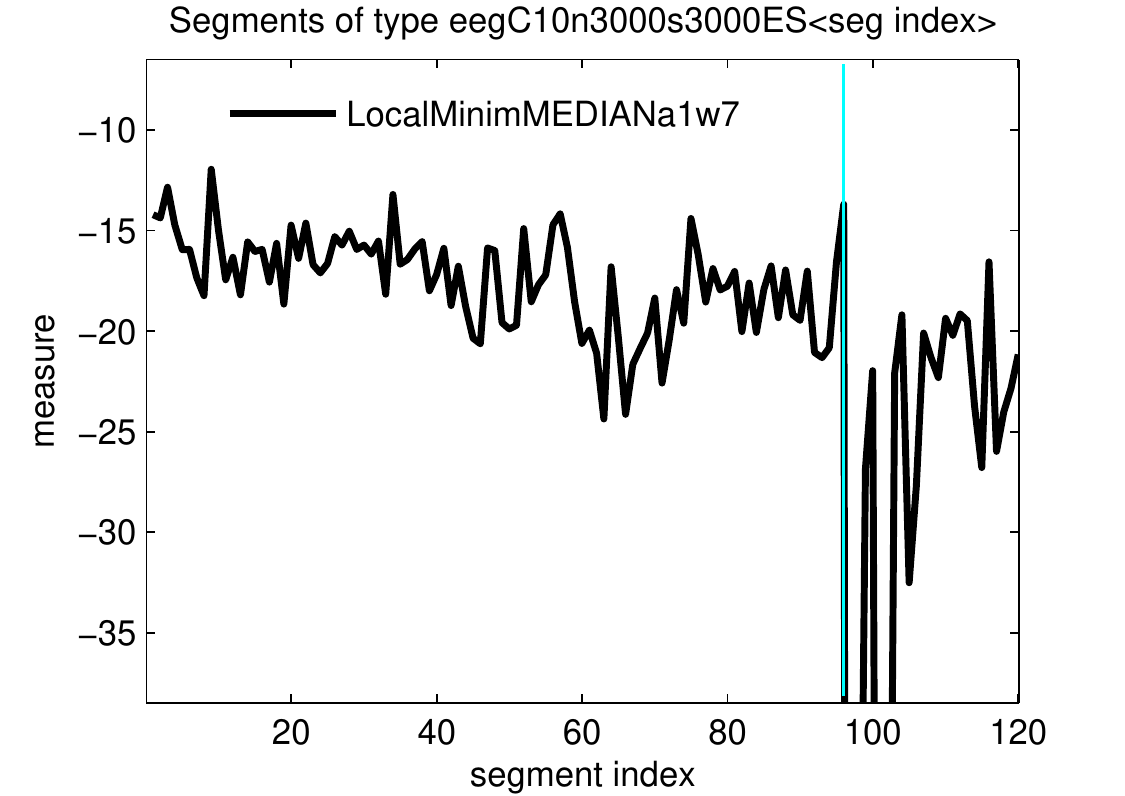}
  \includegraphics[width=7cm,keepaspectratio]{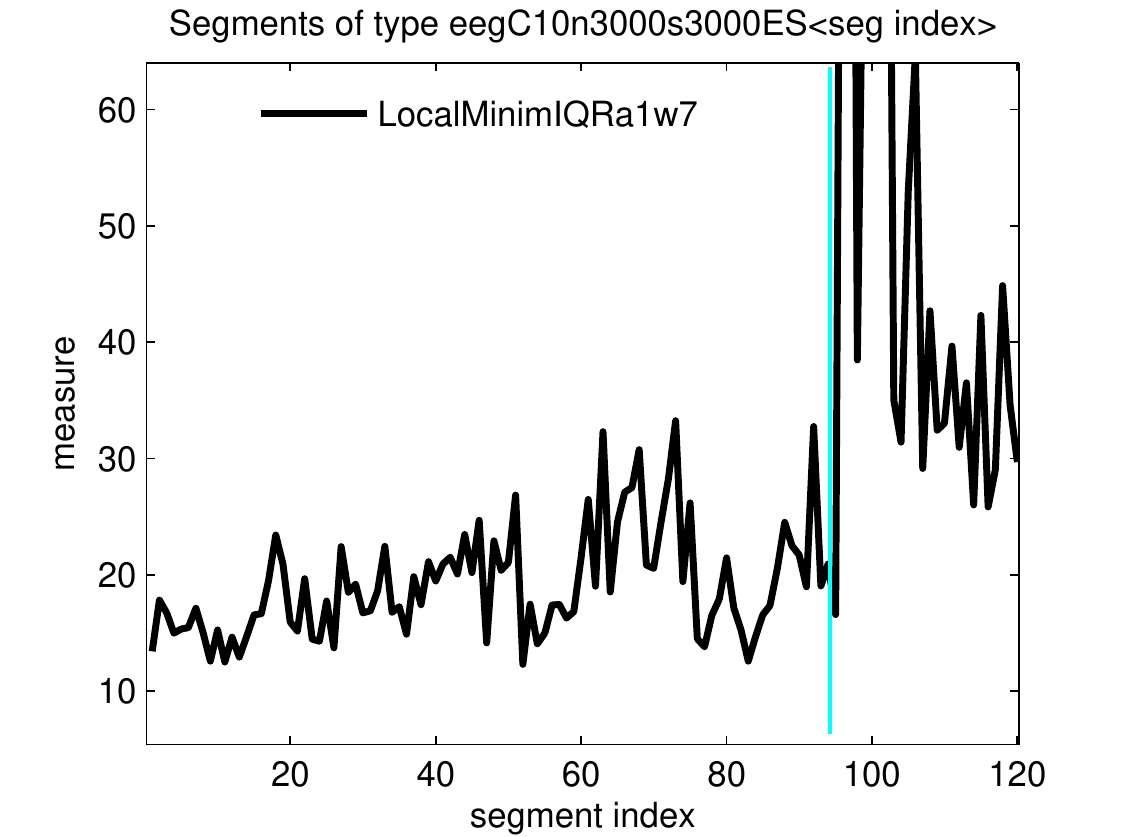}
  }}
\medskip
\centerline{\hbox{
  \includegraphics[width=7cm,keepaspectratio]{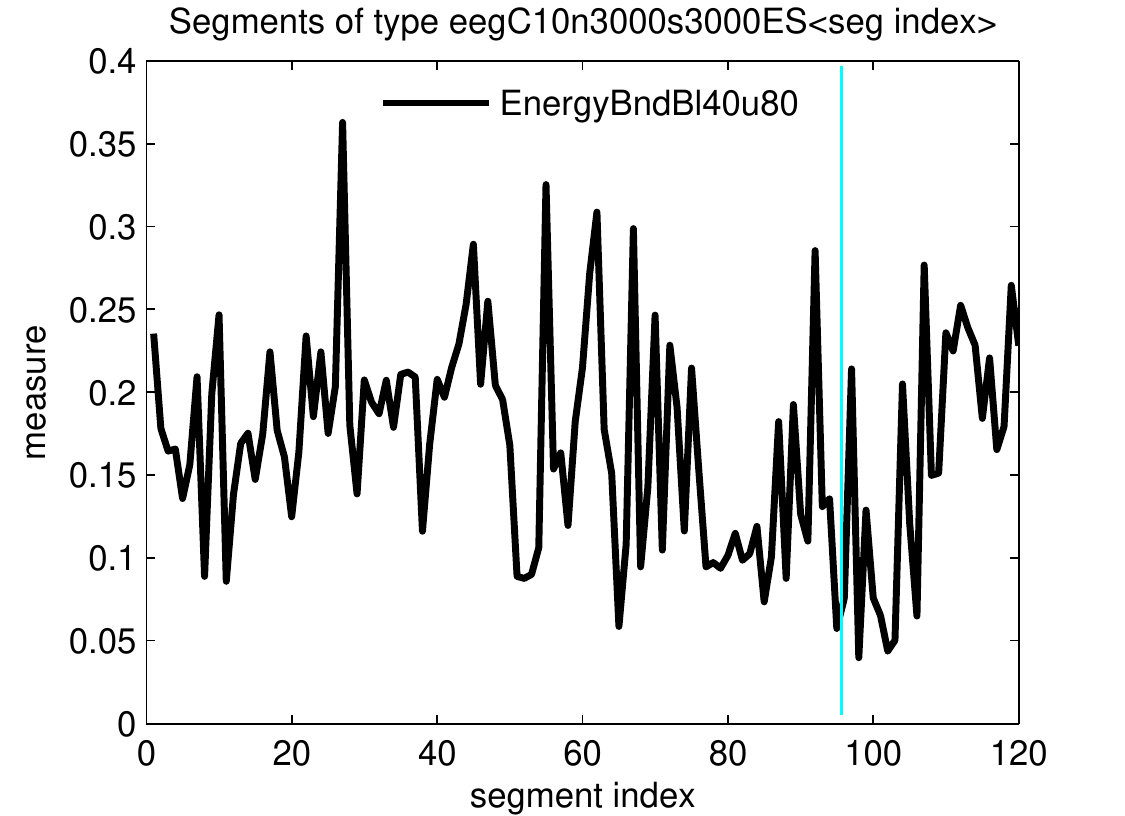}
  \includegraphics[width=7cm,keepaspectratio]{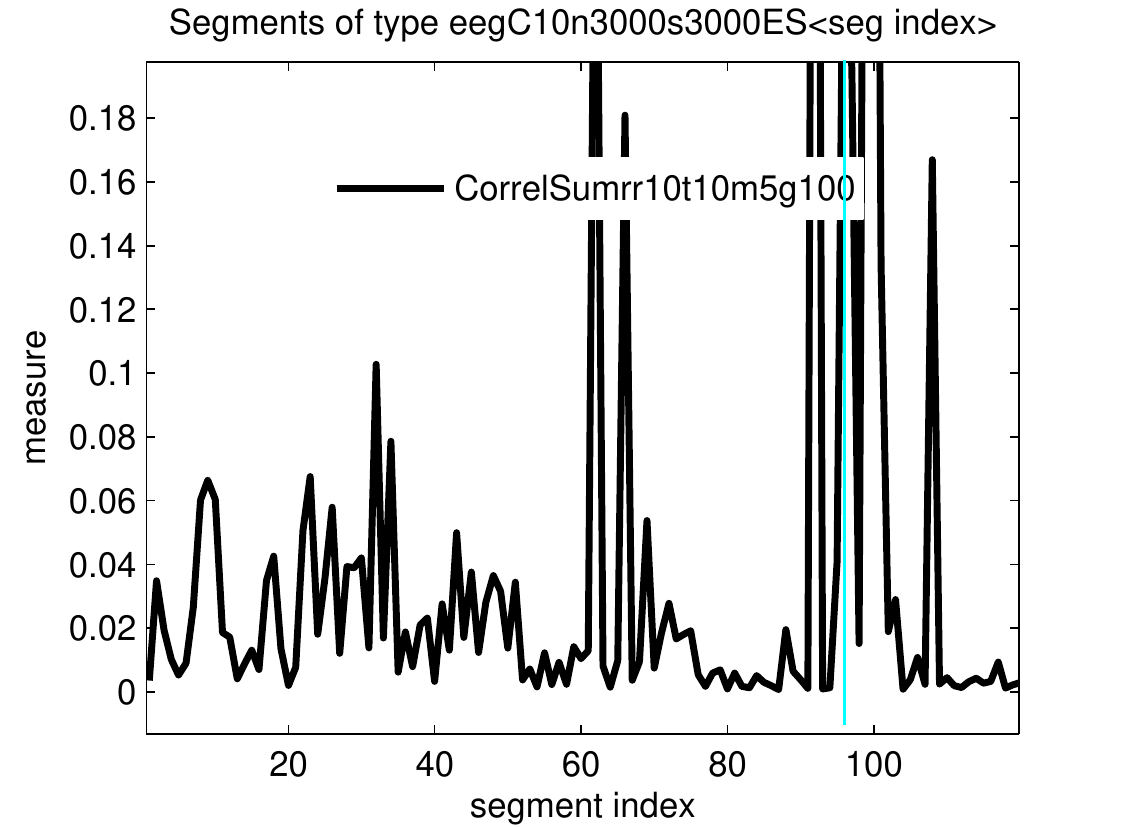}
  }}
\caption{Measure profiles for the segmented time series of channel $10$; top left: median of local minima, top right: IQR of local minima, bottom left: $\theta$ wave, bottom right: correlation sum. Each graph was generated by {\tt View measures - Measure vs segment} and then using the {\tt MATLAB} figure tools the vertical axis was rescaled to ease visualization and a gray (cyan online) vertical line was added at the position of the segment index regarding seizure onset.}
\label{fig:preictalC10mea}
\end{figure}

{\em Step 3. Measure visualization}: To visualize the results on the measures, the {\tt measures vs segments} plot is selected using the facility {\tt View measures}, clicking the corresponding icon in the list of plotting choices (see Figure~\ref{meaviewfig}). A rapid change in amplitude at the time of seizure could be observed in the graphs of all the selected measures. However, a progressive change in the EEG signal could only be observed in the statistics of some features, namely local maxima, local minima and their difference. As shown in Figure~\ref{fig:preictalC10mea}, a downward trend for the median of local minima and upward trend for the IQR of local minima are observed during the pre-ictal period, whereas a less significant downward trend with large fluctuations could be seen in the profile of $\theta$ band.
The profile of correlation sum shows no trend but large fluctuations at the level of the fluctuations during seizure. It is notable that also in the post-ictal period (the time after seizure) the two first measures relax at a different level of magnitude compared to that of the pre-ictal period. This illustration shows that using {\tt MATS} one can easily compare different measures on long records.

\paragraph{Example 2}

The surrogate data test for nonlinearity is often suggested before further analysis with nonlinear tools is to be applied. We perform this test on the EEG segments with index $90$ and $97$ in the previous example regarding the late preictal state (prior to seizure onset) and ictal state (during seizure). Further, we evaluate the performance of three different algorithms that generate surrogate data for the null hypothesis of linear stochastic process underlying the EEG time series, i.e., AAFT, IAAFT and STAP. Note that all three algorithms generate surrogate time series that possess exactly the marginal distribution of the original data, approximate the original linear correlation structure, and are otherwise random, meaning that any nonlinear dynamics that may be present in the original data is removed from the surrogate data. To assess the preservation of the original linear correlations in the surrogate data we use as test statistic the measure of cumulative Pearson autocorrelation for a maximum delay $10$, i.e., sum up the magnitudes of autocorrelation for delays $1$ to $10$. Further, to assess the presence of nonlinear correlations in the original time series that would allow us to reject the null hypothesis, we use as test statistic the measure of cumulative mutual information computed at equiprobable binning for the same maximum delay. A valid rejection of the null hypothesis suggests that the first linear statistic does not discriminate original from surrogate data, whereas the second nonlinear statistic does. The test is performed by {\tt MATS} in the following steps.

{\em Step 1. Data resampling}: We continue the first example and delete from the current time series list all but the segmented time series with index $90$ and $97$. Then using the facility {\tt Resampled time series} we generate $40$ AAFT, IAAFT and STAP surrogates for each of the two EEG time series. For the latter surrogate type, the parameter of degree of polynomial approximation ({\tt pol}) is let to the default value $5$ and the order of autoregressive model ({\tt arm}) is set to $50$ to account for the small sampling time of the EEG data. After completion of the surrogate data generation the current time series list contains $242$ names, the two first are the original data and the rest are the surrogates. For example, the $40$ AAFT surrogates for the segmented EEG time series with index $90$ are {\tt eegC10n3000s3000ES90AAFT1} to {\tt eegC10n3000s3000ES90AAFT40}.

{\em Step 2. Measure selection and computation}: Using the facility {\tt Select / run measures} and {\tt Linear measures - Correlation} the measure of Cumulative Pearson Autocorrelation is selected with the parameter delay ({\tt t}) set to $10$ and then from {\tt Nonlinear measures - Correlation} the measure of Cumulative mutual information with equiprobable bins is selected with the same delay. These two measures are executed on all the $242$ time series in the current time series list and upon completion the current measure list contains the names of the two measures.

\begin{figure}[htb] 
\centerline{\hbox{
  \includegraphics[width=7cm,keepaspectratio]{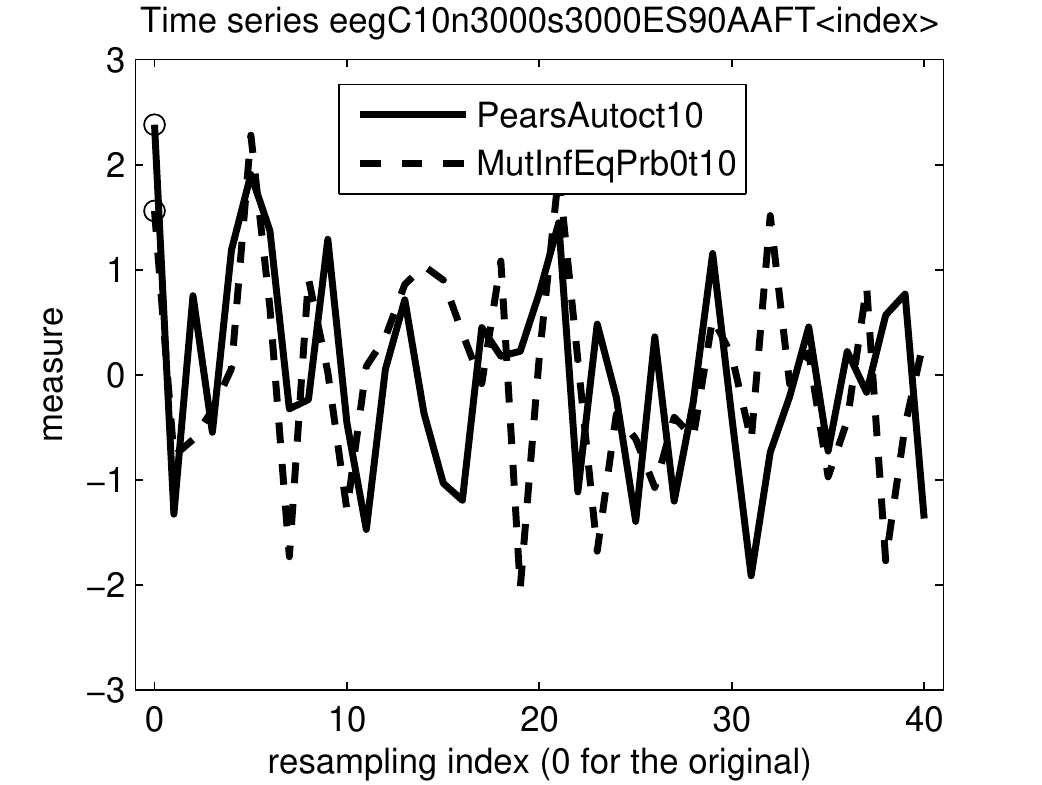}
  \includegraphics[width=7cm,keepaspectratio]{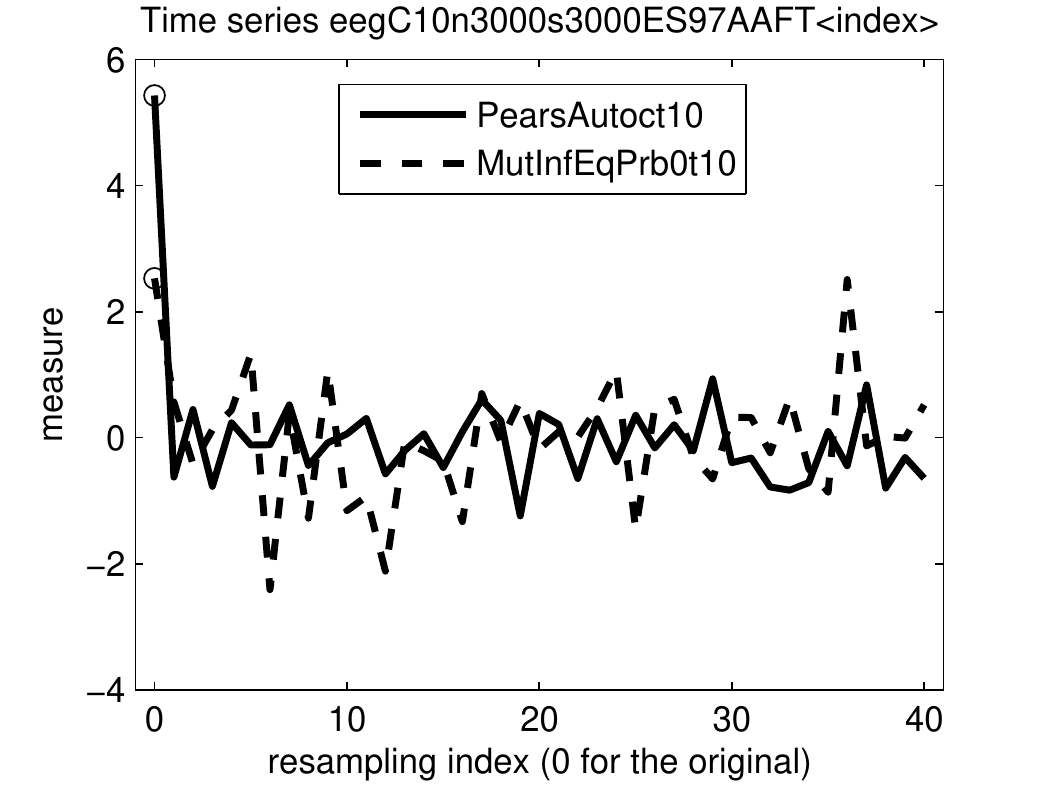}
  }}
\medskip
\centerline{\hbox{
  \includegraphics[width=7cm,keepaspectratio]{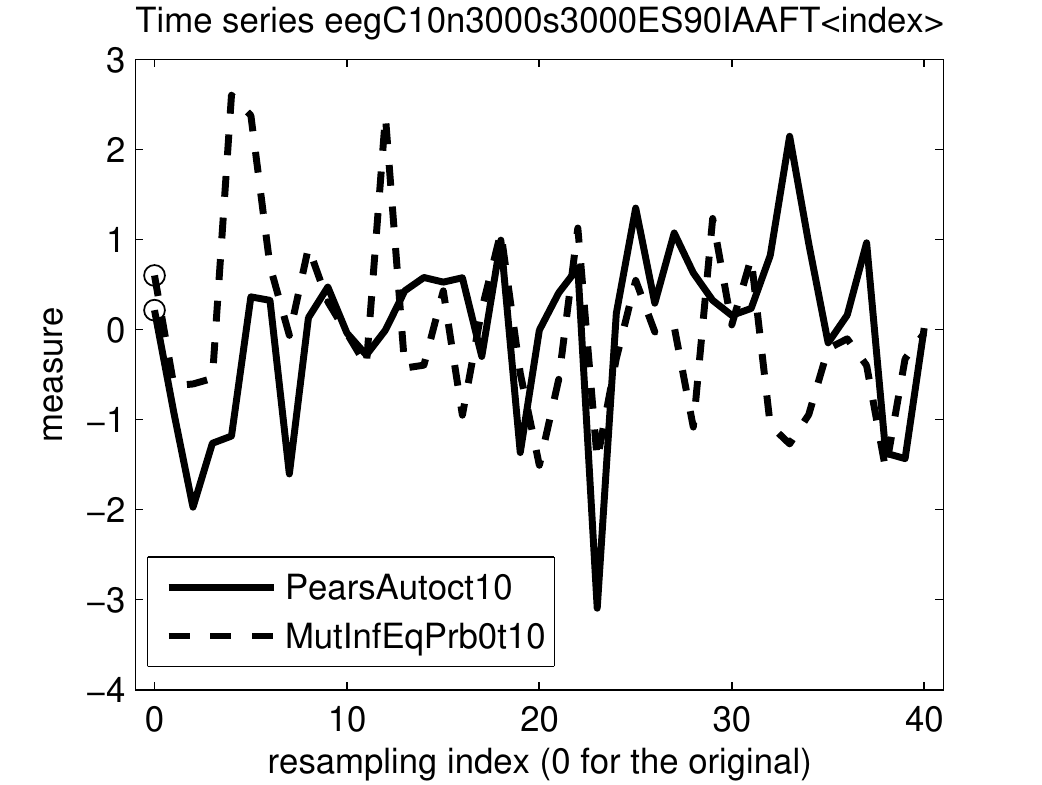}
  \includegraphics[width=7cm,keepaspectratio]{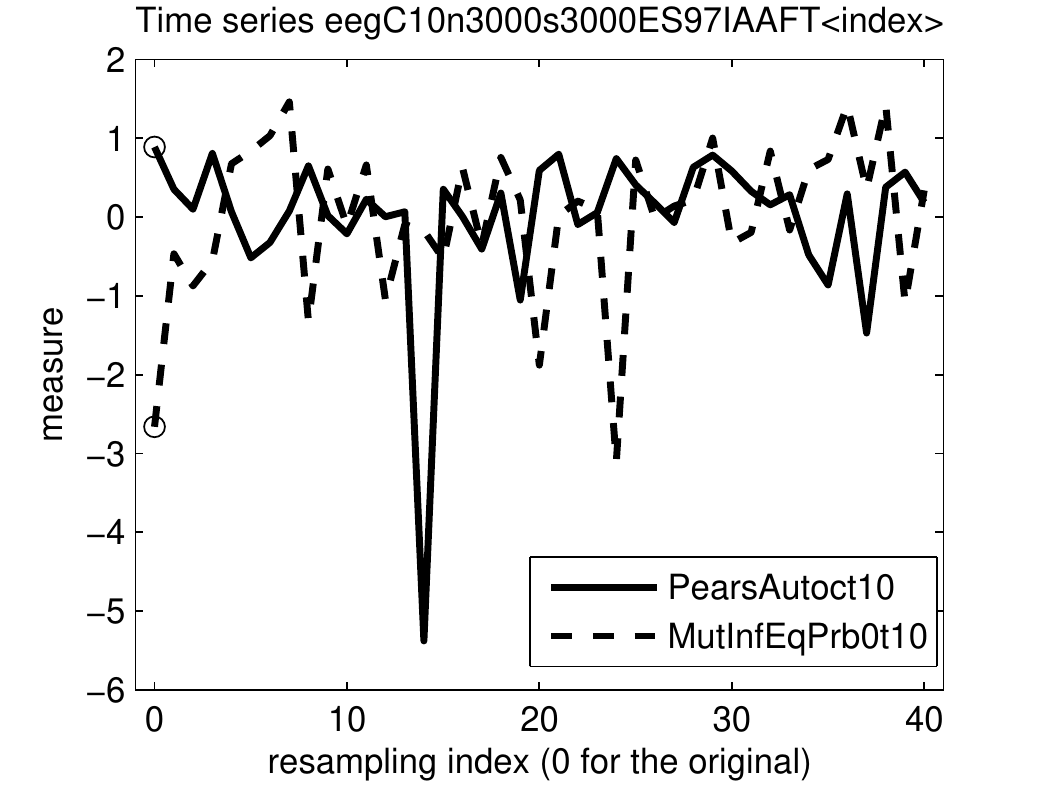}
  }}
\medskip
\centerline{\hbox{
  \includegraphics[width=7cm,keepaspectratio]{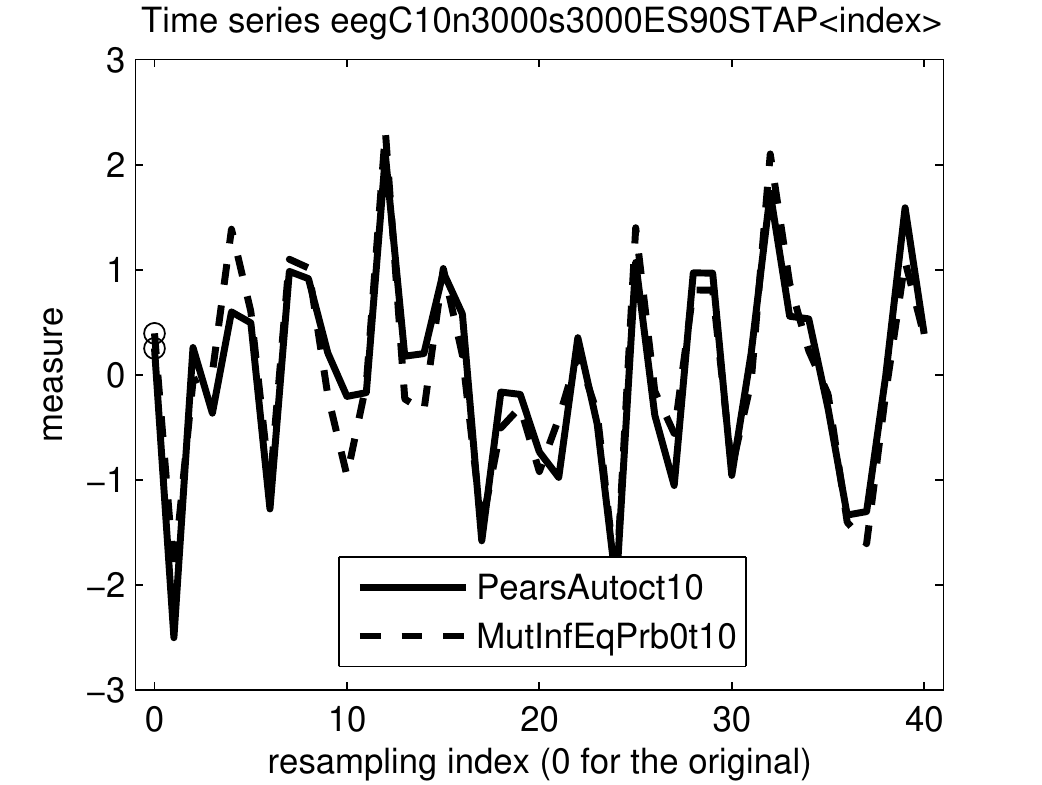}
  \includegraphics[width=7cm,keepaspectratio]{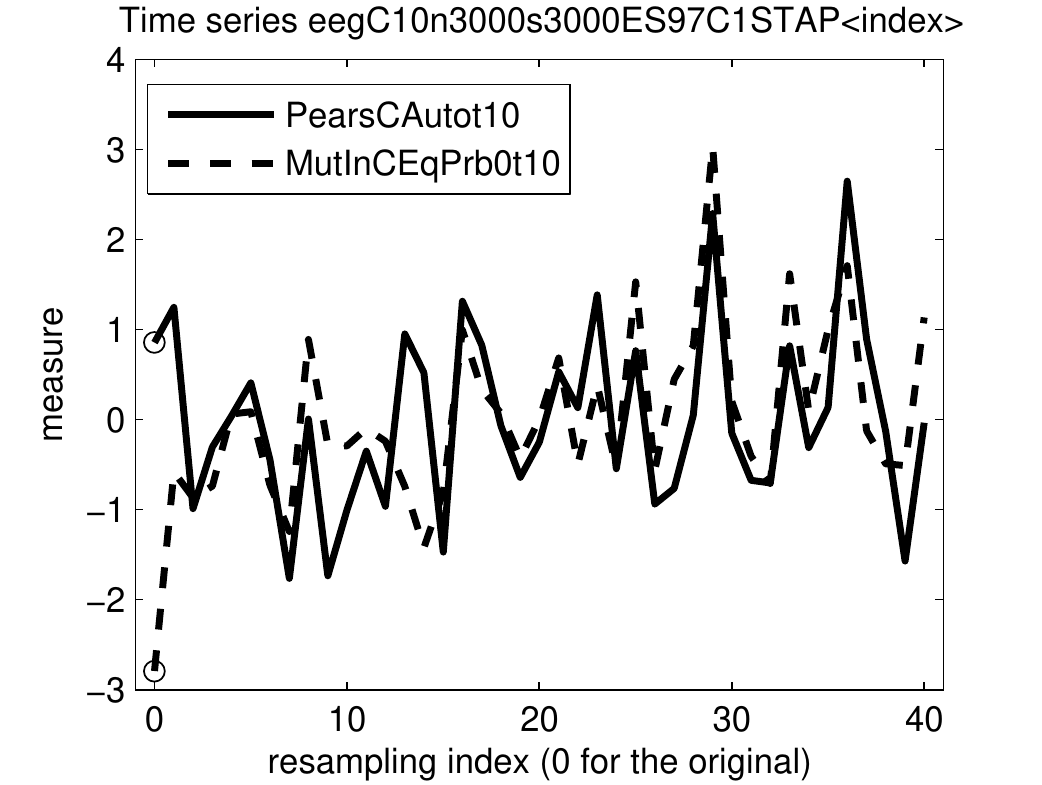}
  }}
\caption{The statistics of cumulative Pearson autocorrelation and cumulative mutual information (equiprobable binning) for maximum delay $10$ for the original time series (index $0$, denoted by open circles) and $40$ resampled time series of three types. For the panels across columns the original time series is the segmented time series with index $90$ (prior to seizure onset) and $97$ (on seizure) and across rows the resampled (surrogate) time series are from the algorithms AAFT, IAAFT and STAP from top to bottom.}
\label{fig:preictalC10sur}
\end{figure}

{\em Step 3. Measure visualization}: {\tt MATS} provides a plot facility under {\tt View Measures} in the main menu for displaying the measure vs resampled time series. In the corresponding GUI, the user is opted to select a surrogate type and then only the time series names matching the selected surrogate type are displayed in the time series list. Further, the user can select the names from this list to be included in the plot for the selected measure. As in any other plot facility, the user can select the graph type, i.e., line, points, or both. In order to allow the same scale for measures of different magnitude range the user can select to normalize the measure values. We have used this facility to make plots of the normalized linear and nonlinear measures for each original time series (with index $90$ and $97$) and surrogate type (AAFT, IAAFT and STAP) and the $6$ plots are shown in Figure~\ref{fig:preictalC10sur}.
The normalized measure value for the original time series is for index $0$ of the x-axis and it is marked with an open circle. We note that AAFT does not preserve the linear autocorrelation of both EEG time series as the first value for index $0$ is above all other $40$ values for the AAFT surrogates (first row of Figure~\ref{fig:preictalC10sur}). The surrogate data test with AAFT is thus invalid and the discrimination with the nonlinear measure for the EEG time series at seizure cannot be taken as legitimate rejection of the null hypothesis since the same holds for the linear statistic. IAAFT does not have the problem of autocorrelation mismatch in the case of pre-ictal state, where no discrimination is observed also with the nonlinear measure, but the problem holds marginally for the ictal state, where no clear discrimination is observed with the nonlinear measure because the value for index 0 is the second smallest (second row of Figure~\ref{fig:preictalC10sur}). The STAP algorithm preserves the linear autocorrelation in both cases, which validates the test, and singles out the original mutual information value only for the EEG time series at ictal state.

\begin{figure}[htb] 
\centerline{\hbox{
  \includegraphics[width=7cm,keepaspectratio]{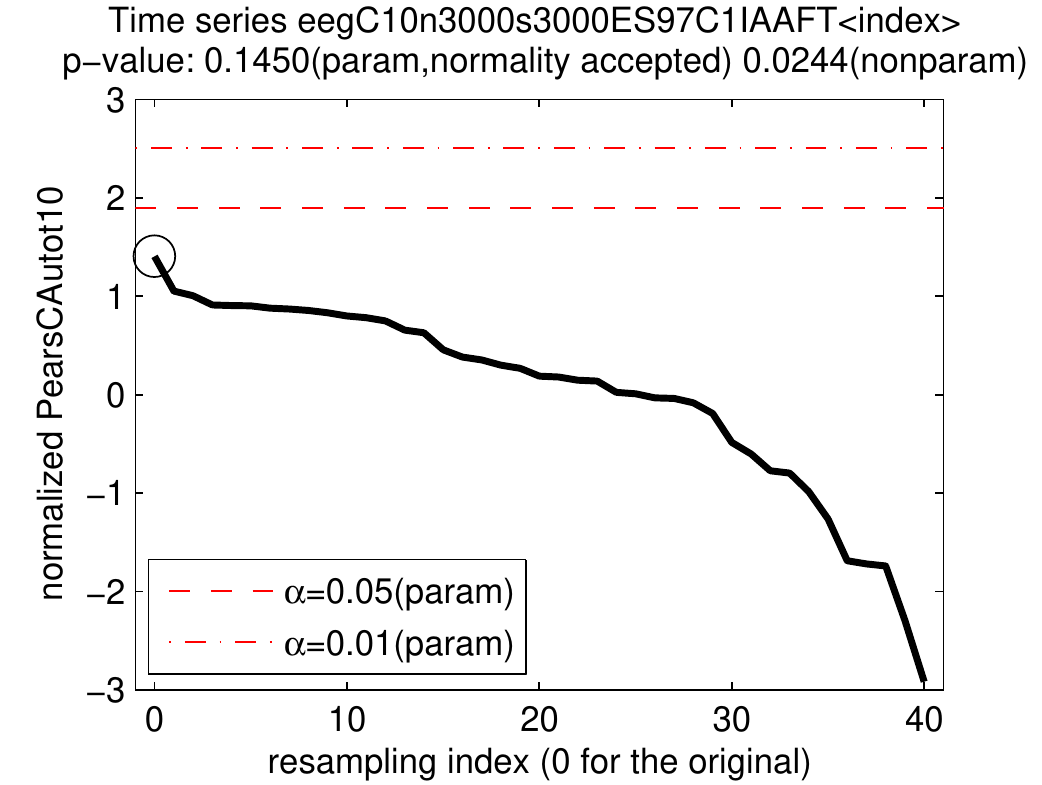}
  \includegraphics[width=7cm,keepaspectratio]{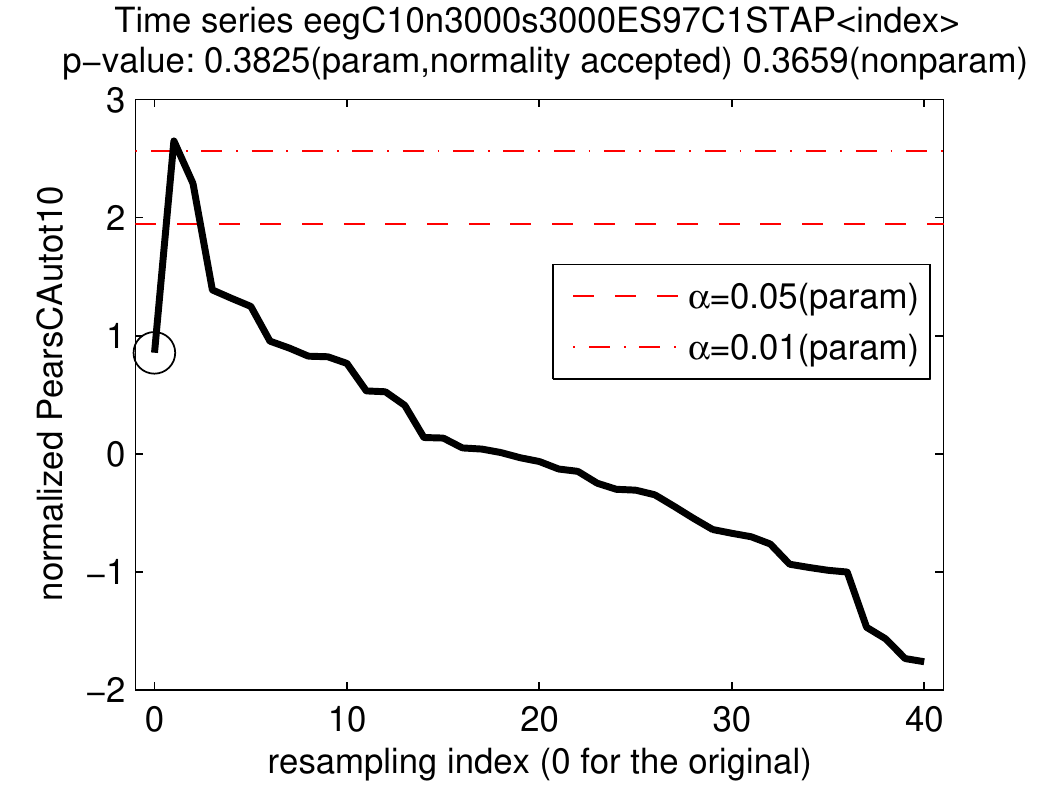}
  }}
\medskip
\centerline{\hbox{
  \includegraphics[width=7cm,keepaspectratio]{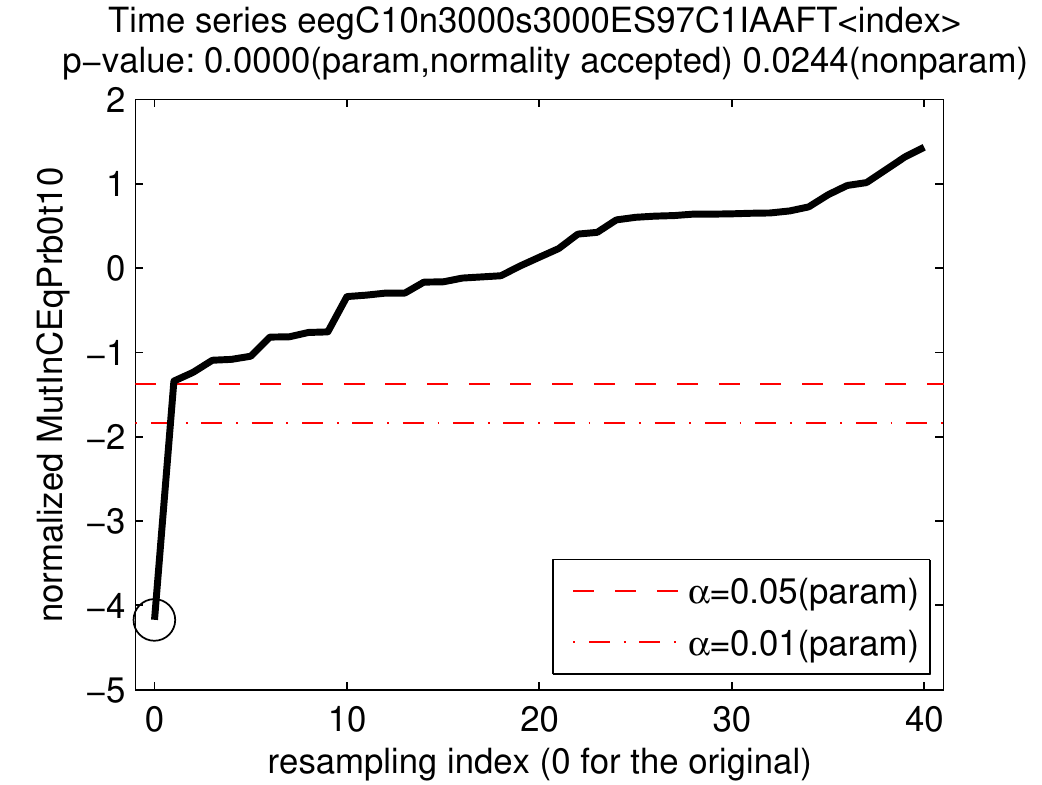}
  \includegraphics[width=7cm,keepaspectratio]{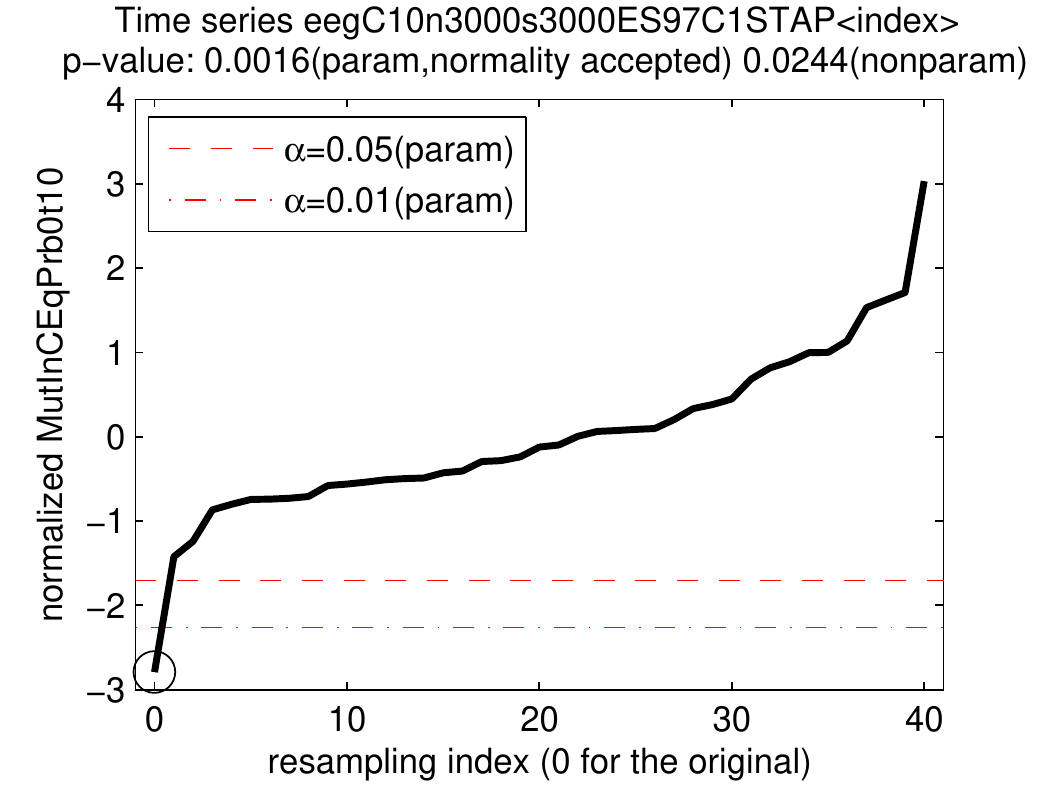}
  }}
\caption{The plots in the first row show the statistics of cumulative Pearson autocorrelation together with the test results. The original time series (given with index $0$) is the segmented EEG time series with index $97$ (on seizure onset) and the $40$ resampled time series are of the IAAFT type (first plot) and STAP type (second plot). The horizontal lines show the significance level for the parametric test at $\alpha=0.01$ and $\alpha=0.05$, as given in the legend. In the second row the same results are shown for cumulative mutual information (equiprobable binning).}
\label{fig:ictalC10IAAFTSTAP}
\end{figure}

The above remarks are based on qualitative assessment from eye-ball judgement of the plots in Figure~\ref{fig:preictalC10sur}. {\tt MATS} provides also quantitative test results in the same visualization facility. The user is opted to display the results of formal tests in the form of $p$-values from the parametric and nonparametric approach (along with a note of acceptance or rejection of the Kolmogorov-Smirnov test for normality of the measure values on the resampled data). Further, horizontal lines at the significance levels of $\alpha=0.01$ and $\alpha=0.05$ are drawn in the plot to allow for visual judgement of the test results. For the case of segment $97$ and IAAFT and STAP surrogates, where the visual inspection is not clearly conclusive, the plots with test results are shown in Figure~\ref{fig:ictalC10IAAFTSTAP}.
It is now shown that for IAAFT, using the linear measure (cumulative Pearson autocorrelation), the parametric test asserts that the measure value on the time series of the segment at the ictal state is within the null distribution, whereas the nonparametric test gives $p=0.0244$ (the smallest $p$-value that can be obtained for a two--sided test from the rank ordering of $41$ values) and concludes for mismatch in the autocorrelation. On the other hand, for STAP the $p$-values for both the parametric and nonparametric approach are large indicating the good match of autocorrelation. For the nonlinear statistic, the null hypothesis of linear stochastic process is rejected for both surrogate types and with both the parametric and nonparametric approach, where the parametric $p$-value is somehow larger for STAP ($p=0.0016$). In all cases, the Kolmogorov-Smirnov test approves the normal distribution of the measure on the surrogate data. The test results with IAAFT and STAP are in agreement to known results in the literature about the drop of complexity of the EEG signal during seizure that allows the detection of low-dimensional nonlinear dynamics.

For illustration purposes the surrogate data test was applied here only to two time series and test statistics, but it is straightforward to include many different measures and time series.

\section{Discussion}
\label{sec:Discussion}

{\tt MATS} is an interactive {\tt MATLAB} toolkit for analysis of scalar time series. The strength of {\tt MATS} is that it can process many time series in one go, which are loaded directly from files or generated using the facilities of segmentation, standardization and resampling. The user can then select among $52$ measures grouped in linear, nonlinear and ``other'' measures, where most of them are defined in terms of a number of parameters. Most of the known measures of analysis of stationary time series are included and the addition of new measures is straightforward by adding a new checkbox for the new measure, with additional boxes and buttons for the parameters, or even a new group of measures in a separate GUI window. The structure of {\tt MATS} is not altered by new insertion of measures as it is built on the current time series list and the current measure list, which are updated dynamically at any selected operation. Thus minimal intervention in the {\tt MATLAB} code of {\tt MATS} is required in order to add measures of choice.

It should be noted that the computation of the selected measures can be slow when a large number of time series are selected or the length of the time series is large (or both). Also, some measures, such as the measure of the local linear model and the measure of the correlation dimension, require long computation time. The {\tt MATLAB} routines are not optimized in every detail, but some effort was made on effective computation, e.g., for data point search a $k$-$D$ tree structure built in {\tt C} is called that is more effective than doing the same structure in {\tt MATLAB} code. Still, there is space for improvement in terms of computation efficiency and it is in the intention of the authors to implement a faster $k$-$D$ tree structure and more efficient calculation of the correlation dimension.

Running on {\tt MATLAB}, {\tt MATS} uses all fine {\tt MATLAB} tools and especially the interactive graphic user interface (GUI) for the selection of the various operations, such as the selection of measures and their parameters, and for visualizing time series and measures in a number of different graph types. {\tt MATS} is meant to be accessible to users with little experience on time series analysis. In addition, a number of checks are performed for each user selection giving out messages for invalid input data and minimizing the risk of stacking at any operation.

{\tt MATS} can be used for various types of analysis requiring minimum user interface. By selecting buttons, checking boxes and specifying parameter values, the user can perform rather complicated tasks. Two such tasks are demonstrated step by step, namely the computation of many measures on consecutive segments of a long time series, and the surrogate data test for nonlinearity on different time series and using different surrogate generating algorithms and test statistics. Further, the user can save results in tables for further processing or later use in {\tt MATS} and make graphs that can also be processed using standard {\tt MATLAB} figure options or stored as files of any desired format.

There are different plug-ins one could think of integrating into this application. We are currently working on a new configuration of this toolkit that includes clustering methods, comparison  of clusters, classifiers and other data mining tools in order to extract information from the computed measures on the time series, an area that is known as feature-based clustering.

\section*{Acknowledgments}

The authors want to thank the two anonymous referees for their constructive remarks on the manuscript and program. The work is part of the research project 03ED748 within the framework of the ``Reinforcement Programme of Human Research Manpower'' (PENED) and it is co-financed at $90\%$ jointly by European Social Fund ($75\%$) and the Greek Ministry of Development ($25\%$) and at $10\%$ by Rikshospitalet, Norway.


\clearpage

\begin{appendix}

\section{Time series names}
\label{AppendixA}

The full name of each time series in the current time series set is comprised of the file name upon loading followed by specific notation for the different operations on the imported time series, as indicated in Table~\ref{tsnames}.
\renewcommand{\baselinestretch}{1}
\begin{table} 
\begin{tabular}{|p{15mm}|p{15mm}|p{20mm}|p{90mm}|c|c|c|c|}
  \hline
  \emph{Character}     & \emph{ordered index}  & \emph{Facility}  & \emph{Description  ($N$=length of time series)} \\
   \hline
  {\tt C} & yes   & Load time series  & The column of the matrix of the loaded file, takes values larger than one if multiple columns are selected\\
    \hline
  {\tt S} & yes   & Segment  & The order of the time series segment in the series of segments when splitting the original time series\\
    \hline
  {\tt n} & yes   & Segment & Length of each segment when segmenting the time series \\
    \hline
  {\tt s} & yes   & Segment & Sliding step in the generation of segments, when $\mbox{{\tt s}}=\mbox{{\tt n}}$ there is no overlapping of segments     \\
    \hline
  {\tt B} & no   & Segment & Residuals are taken from the beginning of the original time series after splitting to segments of length {\tt n} \\
    \hline
  {\tt E} & no   & Segment & Residuals are taken from the end of the original time series after splitting to segments of length {\tt n} \\
    \hline
   {\tt L} & no   & Transform & ``Linear'' standardization resulting in a time series with minimum $0$ and maximum $1$ \\
    \hline
  {\tt N} & no   & Transform & ``Normalized'' or $z$-score standardization resulting in a time series with mean $0$ and standard deviation $1$ \\
    \hline
  {\tt U} & no   & Transform & ``Uniform'' standardization resulting in a time series with uniform marginal distribution \\
    \hline
 {\tt G} & no   & Transform & ``Gaussian'' standardization resulting in a time series with Gaussian marginal distribution\\
    \hline
 {\tt O} & no &	Transform  & ``Log-difference'' transform that first takes the natural logarithm of the data and then the first differences \\
    \hline
 {\tt D} & yes & 	Transform  & ``Detrending'' of the time series with a polynomial fit of given degree. This value follows the code character (the same for the transforms below). \\
    \hline
 {\tt T} & yes & 	Transform & ``Lag difference'' of the time series for a given lag \\
    \hline
 {\tt X} & yes &	Transform & ``Box-Cox'' power transform for a given parameter $\lambda$ \\
    \hline
  {\tt RP} & yes   & Resample & Random permutation. The index of the resampling time series follows the code character (the same for the resampling types below).\\
    \hline
  {\tt FT} & yes   & Resample & Fourier transform \\
    \hline
  {\tt AAFT} & yes   & Resample & Amplitude adjusted Fourier transform \\
    \hline
  {\tt IAAFT} & yes   & Resample & Iterated amplitude adjusted Fourier transform\\
    \hline
  {\tt STAP} & yes   & Resample & Statically transformed autoregressive process\\
    \hline
  {\tt AMRB} & yes   & Resample & Autoregressive model residual bootstrap \\
    \hline

\end{tabular}
  \caption{The time series name notation.}
  \label{tsnames}
\end{table}
Specifically, the code character or string (column one) is possibly followed by an ordered index if the action suggests the generation of multiple time series (specified in column two). The corresponding facility is given in column three and a brief description in column four.

\section{Measure names}
\label{AppendixB}

The full name of each measure in the current measure list is comprised by a $10$ character string of the coded measure name, followed by the measure-specific parameter declaration (one value for each parameter). Each parameter is identified by a character, as shown in the first column in Table~\ref{parnames}.
\begin{table} 
\begin{tabular}{|p{15mm}|p{13mm}|p{22mm}|p{13mm}|p{80mm}|c|c|c|c|c|}
  \hline
  \emph{Character}     & \emph{Multiple values}  & \emph{Valid range}  & \emph{Default} & \emph{Description ($N$=length of time series)}\\
   \hline
    {\tt a} & yes & integers $>0$ & $1$ & moving average filter order, when smoothing the time series for the detection of turning points \\
    \hline
    {\tt b} & yes & integers $>0$ & $0$ & number of bins when partitioning the time series range, if $\mbox{{\tt b}}=0$ then it is determined by the criterion $\mbox{{\tt b}}=\sqrt{N/5}$\\
    \hline
    {\tt c} & yes & [$0$, $1$] & $0.5$ & correlation sum, given as input in order to find the corresponding radius \\
    \hline
    {\tt e} & no & integer $\ge2$ & $10$ & escape factor, for assessing the false nearest neighbors \\
    \hline
    {\tt f} & no & [$0.1$, $0.9$]& $0.5$ & fraction for the test set, e.g., if $\mbox{{\tt f}}=0.3$ the length of the test set is $0.3N$ \\
    \hline
    {\tt g} & no & integer $\ge0$ & $0$ & Theiler's window, to exclude temporally correlated points from the set of neighboring points\\
    \hline
    {\tt h} & yes & integers $>0$ & $1$ & lead time for fitting and prediction\\
    \hline
    {\tt k} & yes & integers $>0$ & $1$ & number of nearest neighbors\\
    \hline
    {\tt l} & no & [$0$, $0.5$] & multiple cases & lower band frequency, for the computation of energy in frequency bands \\
    \hline
    {\tt m}  & yes & integers $>0$ & $1$ & embedding dimension or AR model order \\
    \hline
    {\tt o} & no & integer$\,\ge10$ & $100$ & number of radii to calculate the correlation sum and estimate the correlation dimension\\
    \hline
    {\tt p} &	yes & integers $>0$ & $1$ & order of the moving average part in ARMA \\
    \hline
    {\tt q} & no & integer $\ge0$ & $0$ & truncation parameter, $\mbox{{\tt q}}\!\!=\!\!0$ for local average map, $\mbox{{\tt q}}\!\!=\!\!\mbox{{\tt m}}$ for OLS solution of the local linear model parameters, $\mbox{{\tt q}}\!\!<\!\!\mbox{{\tt m}}$ for PCR solution \\
    \hline
    {\tt r} & yes & [$0$, $1$] & $0.1$ & radius for the computation of the correlation sum (the time series is standardized in $[0,1]$) \\
    \hline
    {\tt s} & yes & integers $>0$ & $4$ & size of scaling window, i.e., $\mbox{{\tt s}}=r_2/r_1$, where $r_2-r_1$ is the length of the scaling window  \\
    \hline
    {\tt t} & yes & integers $>0$  & $1$ & delay time\\
    \hline
    {\tt u} & no & [$0$, $0.5$] & multiple cases & upper band frequency, for the computation of energy in frequency bands; $\mbox{{\tt u}}>\mbox{{\tt l}}$ should hold  \\
    \hline
    {\tt w} & yes & integers $>0$ & $1$ & offset for the local window of length $2\mbox{{\tt w}}+1$, used in the detection of turning points\\
  \hline
 \end{tabular}
  \caption{Parameter name notation.}
  \label{parnames}
\end{table}
The parameters may bear only a single value or a range of values given in standard {\tt MATLAB} syntax, e.g., {\tt 1:10} or {\tt [1 5:10 20]} (see second column) and may be constrained to specific ranges (see third column). Default values for the parameters are given in column four and a brief description follows in column $5$. Note that if a range of values is given for a parameter, a different measure name for each parameter value is listed in the current measure list.

\end{appendix}

\end{document}